\documentclass[twocolumn]{aastex631}

\usepackage{graphicx}
\usepackage{rotating}

\begin{document}

% \title{Optimizing Habitable Worlds Observatory's Orbital Characterization of Cold Giants and Habitable Worlds}
\title{Requirements for Joint Orbital Characterization of Cold Giants and Habitable Worlds with Habitable Worlds Observatory}

\author[0000-0002-6650-3829]{Sabina Sagynbayeva}
\affiliation{Department of Physics and Astronomy, Stony Brook University, Stony Brook, NY 11794, USA}
\affiliation{Kavli Institute for Theoretical Physics, University of California, Santa Barbara, CA 93106, USA}

\author[0009-0006-4626-832X]{Asif Abbas}
\affiliation{Department of Astronomy, New Mexico State University, 1320 Frenger Mall, Las Cruces, NM 88003, USA}

\author[0000-0002-7084-0529]{Stephen R. Kane}
\affiliation{Department of Earth and Planetary Sciences, University of California, Riverside, CA 92521, USA}

\author[0000-0001-6975-9056]{Eric L. Nielsen}
\affiliation{Department of Astronomy, New Mexico State University, 1320 Frenger Mall, Las Cruces, NM 88003, USA}

\author[0000-0001-5684-4593]{William Thompson}
\affiliation{NRC Herzberg Astronomy and Astrophysics,
5071 West Saanich Road,
Victoria, BC, V9E 2E7, Canada}

\author[0000-0002-3199-2888]{Sarah Blunt}
\affiliation{Department of Astronomy \& Astrophysics, University of California, Santa Cruz, CA 95064, USA}

\author[0000-0002-7670-670X]{Malena Rice}
\affiliation{Department of Astronomy, Yale University, 219 Prospect St., New Haven, CT 06511, USA}

\author[0000-0002-8035-4778]{Jessie L.\ Christiansen}
\affiliation{NASA Exoplanet Science Institute, IPAC, MS 100-22, Caltech, 1200 E.\ California Blvd, Pasadena, CA 91125}

\author[0000-0001-5737-1687]{Caleb K.\ Harada}
\altaffiliation{NSF Graduate Research Fellow}
\affiliation{Department of Astronomy, 501 Campbell Hall \#3411, University of California, Berkeley, CA 94720, USA}

% \author{Courtney Dressing}
% \affiliation{Berkeley}

\author[0000-0003-4150-841X]{Elisabeth R. Newton}
\affiliation{Department of Physics and Astronomy, Dartmouth College, Hanover NH 03755, USA}

\author[0000-0002-9017-3663]{Yasuhiro Hasegawa}
\affiliation{Jet Propulsion Laboratory, California Institute of Technology, Pasadena, CA 91109, USA}

\author[0000-0001-5032-1396]{Philip J. Armitage}
\affiliation{Department of Physics and Astronomy, Stony Brook University, Stony Brook, NY 11794, USA}
\affiliation{Center for Computational Astrophysics, Flatiron Institute, 162 Fifth Avenue, New York, NY 10010, USA}

\author[0000-0002-6939-9211]{Tansu Daylan}
\affiliation{Department of Physics and McDonnell Center for the Space Sciences, Washington University, St. Louis, MO 63130, USA}

% \author{Romy Rodriguez}
% \affiliation{CfA}




\begin{abstract}
We determine optimal requirements for the joint detection of habitable-zone planets and cold giant planets with the Habitable Worlds Observatory (HWO). Analysis of 164 nearby stars shows that a coronagraph outer working angle (OWA) of 1440 milliarcseconds (mas) is necessary to achieve 80-90\% visibility of cold giants. Approximately 40 precursor radial velocity measurements with 1 m/s precision are required to adequately constrain orbital parameters before HWO observations. We demonstrate that 6-8 astrometric measurements distributed across the mission timeline, compared to radial velocity constraints alone and to astrometry constraints alone, significantly improve orbital parameter precision, enabling direct determination of orbital inclination with uncertainties of $\pm0.8-3^\circ$. For habitable-zone planet characterization, 4-5 epochs provide moderate confidence, while high-confidence (95\%) confirmation requires 8+ observations. These specifications are essential for the comprehensive characterization of planetary system architectures and understanding the potential habitability of terrestrial exoplanets.

\end{abstract}

% \keywords{planets and satellites: detection --- Ultraviolet astronomy(1736) --- History of astronomy(1868) --- Interdisciplinary astronomy(804)}


\section{Introduction} \label{sec:intro}

The search for habitable worlds beyond our solar system stands at a pivotal moment with the planned Habitable Worlds Observatory (HWO) mission, which aims to characterize approximately 25 terrestrial exoplanets, where characterization includes both determining orbital parameters and obtaining direct imaging spectroscopy to assess atmospheric composition — potentially including biosignature gases. Although habitable conditions may be present in a variety of environments on planets and their satellites, rocky extrasolar planets whose surfaces can sustain the presence of liquid water over long time scales represent the best-understood and most Earth-like possibility \citep[e.g.,][]{Rothschild2001, Schulze2006}.
%This ambitious goal reflects the scientific quest to detect and characterize terrestrial planets that may sustain the presence of surface liquid water, potentially producing conditions amenable to long-term habitability. 
HWO will thus target nearby stars whose habitable zones (HZ) fall within the angular separation range set by the coronagraph’s inner and outer working angles \citep{Kasting1993,kane2012a,kopparapu2013a,kopparapu2014,hill2023}.

A significant challenge for the HWO mission is that direct imaging has not yet successfully identified HZ planets. Our current understanding of HZ planet demographics largely relies on other techniques such as transits, radial velocity (RV), microlensing, and, in the near future, astrometry from Gaia.
Those demographic studies show that close-in systems of super-Earth and sub-Neptune class planets are extremely abundant, while the frequency of cold giant planets is lower and dependent on stellar host type. 
Higher occurrence rates are inferred for sub-Neptunes around M dwarfs from microlensing surveys \citep{Suzuki2016}, and lower rates for giant planets around Sun-like stars \citep{Fulton2021,Fernandes2019}. 

The abundance of cold giants around stars hosting rocky planets is not fully known, although recent studies \citep[e.g.,][]{Zhu2019, Bryan2019, Bryan2024, Bryan2025, Bonomo2023, Weiss2024, zandt2025} have explored the conditional probabilities of finding both close-in Earth-like planets and cold Jupiters. These studies present a variety of results about whether cold giant planets are more, less, or equally frequent given the presence of sub-Neptunes, as well as stellar type and orbital configuration. Nonetheless, full characterization of HZ planets and their planetary systems is critical to optimizing HWO's mission design. Failing to measure key attributes, such as the presence of giant planets near the snowline, could lead to suboptimal decisions that compromise mission success. The architecture of our own solar system, with Jupiter’s dynamical influence on inner planets, serves as an important example of how such giant planets can affect the occurrence of rocky HZ planets \citep[e.g.,][]{Raymond2004, kane2024a}. At the most basic level, giant planets can significantly destabilize the orbits of HZ terrestrial planets within a system \citep{barnes2004b,veras06,kopparapu2010,raymond11,kane2015b,kane2019e,kane2023c,kane2023d}. Moreover, the habitability of terrestrial planets is inextricably linked to their volatile content and, in particular, their water abundance, which, in principle, can vary by several orders of magnitude depending on the formation history of the planetary system. Although planets can acquire water from {\em in situ} and gas disk processes \citep{young23,kral24}, the most well-studied mechanism for water delivery to the Earth relies on impacts from water-rich bodies that formed beyond the snow line \citep{morbidelli00, Meech2019, Raymond2017}. Giant planets play an essential dynamical role in this story \citep{Raymond2004,kane2024a}. These considerations motivate the need for pre-cursor radial velocity (RV) observations \citep{kane2024e} to characterize the orbits of any cold giants as part of HWO target selection \citep{kane2024d}, along with studies of how HWO itself, and future facilities operating concurrently with HWO, can achieve a comprehensive understanding of the full range of orbital and atmospheric parameters of each observed planetary system. 
%which is fundamentally shaped by planetary system architecture and formation history \citep{Raymond2017}. Giant planets play a crucial role in this story by directing volatiles from beyond the snow line to the inner system \citep{Raymond2004,kane2024a}. While the frequency of cold giants appears to vary with stellar host type, with higher occurrence rates around M dwarfs detected through microlensing surveys \citep{Fernandes2019,Suzuki2016}, and lower rates around Sun-like stars \citep{Fulton2021,Fernandes2019}, these planets can dramatically alter the distribution and transport of water-rich materials throughout their planetary systems. Furthermore, the presence of giant planets can significantly destabilize the orbits of HZ terrestrial planets within a system \citep{barnes2004b,kopparapu2010,kane2015b,kane2019e,kane2023c,kane2023d}.
%This motivates the need for pre-cursor radial velocity (RV) observations \citep{kane2024e} to characterize the orbits of any cold giants as part of HWO target selection \citep{kane2024d}.

The Astronomy \& Astrophysics Decadal Survey  \citep[Astro2020;][]{NAP26141} recommended the development of a telescope whose capabilities are defined by the science goal of measuring potential biomarkers in the atmospheres of 
roughly 25 HZ planets. The proposed observatory, now known as HWO, is notionally envisaged as a roughly 6m inscribed diameter space telescope with wide wavelength coverage from the ultraviolet to the infrared, equipped with a next-generation coronagraph. The details of the telescope architecture, however, have yet to be decided, and it is important to assess how different possible choices impact the range of ancillary science that HWO could perform. Here, we focus on understanding the observational requirements needed for HWO to measure the orbital parameters of HZ planets {\em and} cold Jupiters as part of the same observing campaign. As discussed above, the ability to detect both HZ candidates and their outer companions (or lack thereof) around nearby stars would help answer fundamental questions about planetary system architecture and its role in creating habitable conditions.

%Currently, several working groups are actively developing different aspects of the telescope architecture, \textit{which is not yet known}. While the primary mission objective centers on detecting and characterizing HZ planets, the telescope's capabilities will enable the detection of cold Jupiters, offering valuable insights into the architectural diversity of planetary systems. We focus on understanding the observational requirements needed to achieve the mission's scientific goals, particularly the orbital parameters of HZ planets and cold Jupiters, as a valuable bonus science case. We also note that HWO's capability to detect gas giants at wider separations during the same observing campaign as the detection of HZ planets is particularly valuable because the presence or absence of cold Jupiters may significantly influence the habitability of inner rocky planets. The ability to detect both HZ candidates and their outer companions (or lack thereof) around nearby stars will help answer fundamental questions about planetary system architecture and its role in creating habitable conditions.

\begin{figure}[hbt!]
    \center
    \includegraphics[width=0.46\textwidth]{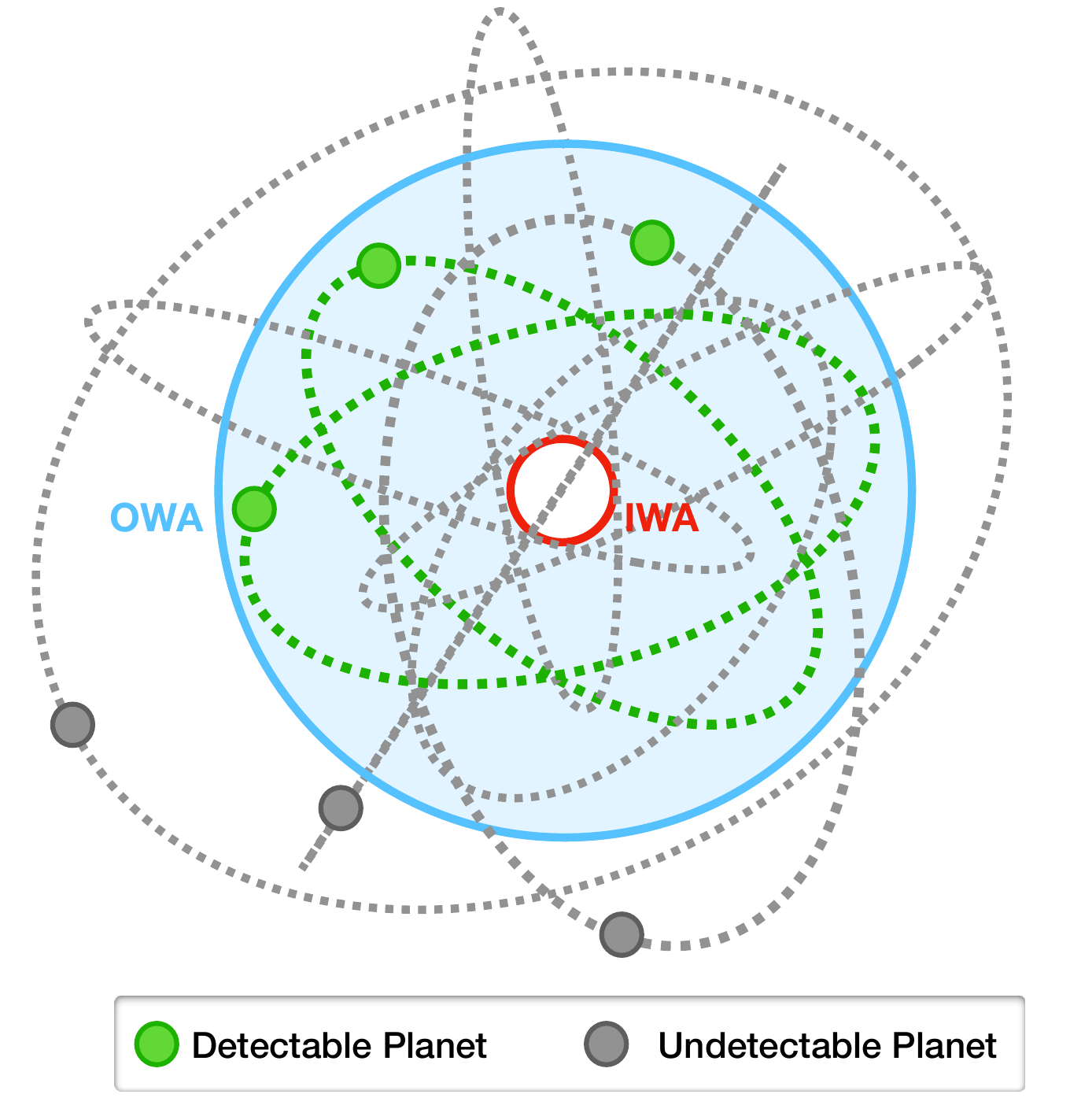}
        \caption{Schematic representation of coronagraph working angles. The Inner Working Angle (IWA) and Outer Working Angle (OWA) define the detectable region (dotted pattern) where exoplanets can be observed. The central star is blocked by the coronagraph, while planets can be detected between the IWA (red dashed line) and OWA (blue dashed line). Green dots represent detectable planets, while gray dots indicate planets that fall outside the observable region.}
        \label{fig:coronagraph}
\end{figure}

In this paper, we examine how the coronagraph's inner working angle (IWA) and outer working angle (OWA) constrain our ability to detect HZ planets \textit{and} cold Jupiters residing in the same planetary system. These working angles represent fundamental observational constraints that define the spatial region where direct imaging techniques can effectively detect and characterize exoplanets (see Figure \ref{fig:coronagraph}). The IWA is the smallest angular separation from the host star at which a companion object can be reliably detected, primarily limited by diffraction effects and typically approximated as $\lambda/D$, where $\lambda$ is the observing wavelength and $D$ is the telescope diameter. The OWA represents the maximum observable angular separation, constrained by the detector's field of view and the size of the dark hole, set by the Nyquist limit of the deformable mirror. We aim to constrain the orbital characteristics of planets that HWO can potentially observe with a given coronagraph architecture, informing both target selection and telescope design requirements. 

The remainder of this paper is organized as follows. Section \ref{sec:telescope-arch} describes the proposed telescope architecture and its capabilities for detecting cold giants. Section \ref{section: coldjupiters} presents our analysis and results on inferring the orbits of cold Jupiters, with \ref{subsection: methodology2} detailing our methodology. Section \ref{section:hzplanets} describes our analysis of constraining the orbits of HZ planets. Section \ref{sec:discussion} discusses the implications of our findings for target selection and mission design. Finally, Section \ref{sec:conclusions} summarizes our conclusions and their implications for the HWO mission.

\section{The Catalog} \label{sec:catalog}

\begin{figure*}[hbt!]
    \center
    \includegraphics[width=0.9\textwidth]{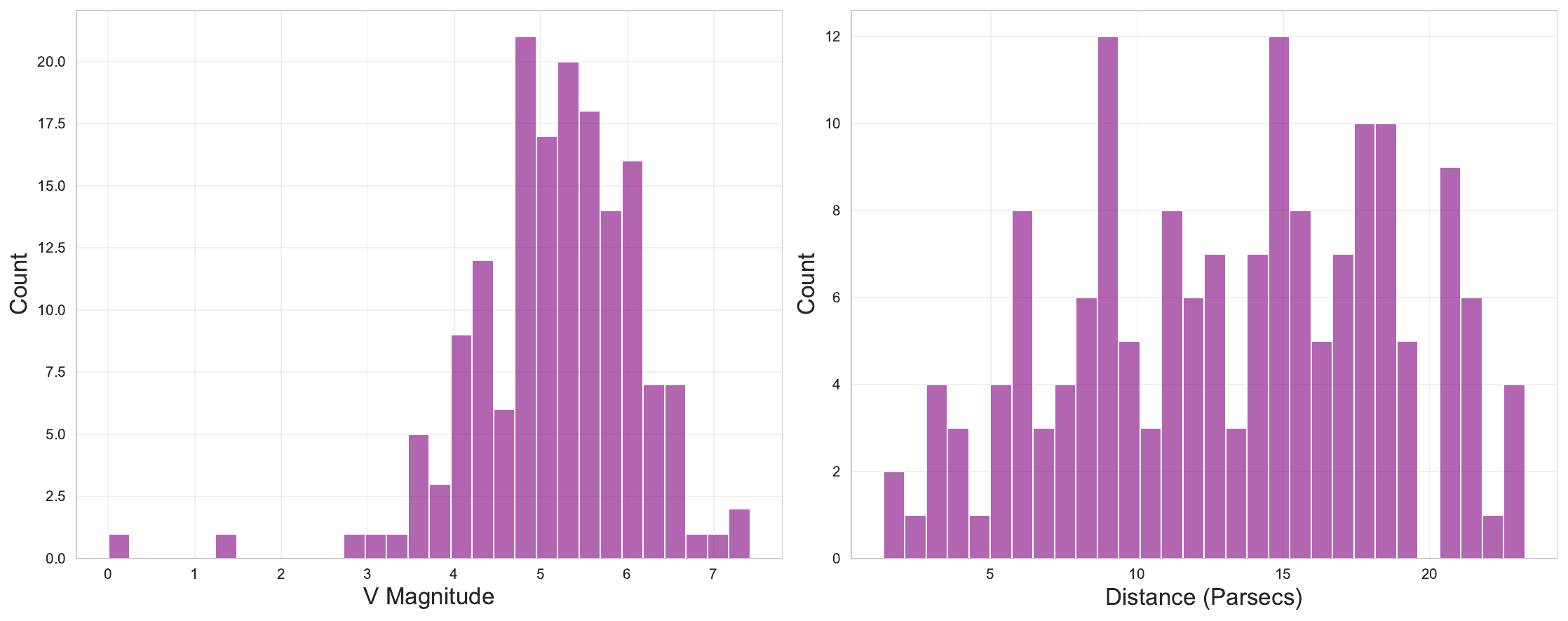}
        \caption{Properties of stars in the ExEP catalog. Left: Distribution of V magnitudes, showing a peak between V = 5-6, indicating a sample dominated by bright stars. Right: Distribution of distances, showing a relatively uniform spread of targets out to 20 parsecs, with most stars between 5-20 parsecs, making this a sample of nearby stars.}
        \label{fig:dist-mag}
\end{figure*}

\cite{Mamajek2024} created a preliminary target list for the Habitable Worlds Observatory's search for potentially habitable planets around nearby stars, \textit{the NASA Exoplanet Exploration Program’s Mission Star List for the Habitable Worlds Observatory} \citep[hereinafter, ExEP catalog;][]{Mamajek2024}. The list includes 164 stars within 25 parsecs of Earth (see Figure \ref{fig:dist-mag}), mostly consisting of stars similar to the Sun: 66 F-type stars, 55 G-type stars, 40 K-type stars, and 3 M-type stars. Figure \ref{fig:dist-mag} shows the distributions of V magnitudes and distances for the ExEP catalog. Most stars in the sample are bright, with $4<V<7$, peaking around V = 5-6. The distance distribution shows that the HWO targets are some of the nearest neighbors in the local solar neighborhood, extending to 25 parsecs with a relatively uniform spread between 5 and 20 parsecs. 

Our analysis consisted of two complementary approaches: (1) a study of angular separations across the ExEP stellar catalog \citep{Mamajek2024} to optimize HWO's working angle specifications, and (2) an orbital inference study comparing RV-only, Astrometry-only versus RV+astrometry constraints on geometric orbital parameters. 

\subsection{Why nearby stars?}

The nearby bright star population within 25 parsecs provides an optimal sample for HWO's science objectives. At these distances, planets at 1 AU would have angular separations of $\sim40$ milliarcseconds from their host stars ($1 \rm AU/25 pc = 40 mas$). While an 8m telescope operating at 1-micron wavelength with an idealized coronagraph (1$\lambda/D$ inner working angle) would be limited to separations greater than 26 mas, various HWO architecture options can accommodate this range. For HWO's goal of characterizing approximately 25 Earth-sized planets, target selection must balance these distance limitations with apparent magnitude constraints (generally $V < 8$ for Earth-analog searches) and factors affecting coronagraph performance \citep{Tuchow2024}.

These nearby stars are simultaneously accessible to precise RV measurements, high-precision astrometry, transit searches, and direct imaging, enabling comprehensive characterization of planetary systems. HWO's capability to detect both habitable zone candidates and gas giants at wider separations during the same observing campaign is particularly valuable, as the presence or absence of cold Jupiters may significantly influence the habitability of inner rocky planets. This multi-technique approach will help answer fundamental questions about planetary system architecture and its role in creating habitable conditions. 

\section{Notes on Proposed Telescope Architecture}\label{sec:telescope-arch}
The detection and characterization of exoplanets with HWO relies fundamentally on the telescope's ability to suppress starlight and isolate planetary signals. The technique of direct imaging requires suppressing the light from host stars with optical devices such as coronagraphs, enabling the faint planetary point source to be distinguished from stellar glare.

The primary detection criteria are determined by three key instrument parameters that shape the observable parameter space. First, the inner working angle (IWA) defines the minimum angular separation at which planets can be detected. For a coronagraph system, this is given by:
\begin{equation}\label{eq:iwa}
     IWA = N \times \lambda/D, \\
\end{equation}
where $N$ is typically between 2-4, $\lambda$ is the observing wavelength, and $D$ is the telescope diameter. For HWO's (assumed) 6m primary mirror operating at $\lambda = 1$ $\mu m$, this yields an IWA ranging from approximately 25-50 milliarcseconds (mas). For reference, at a distance of 20 pc, this corresponds to physical separations of 0.5-1 AU; the detection capabilities thus scale linearly with stellar distance.

The outer working angle (OWA) determines the maximum detectable separation:
\begin{equation}\label{eq:owa}
     OWA = X \times \lambda/D, \\
\end{equation}
%                                   (2)
where $X$ is determined by the specific coronagraph design. 
% For HWO, X is expected to range from 10-15 based on current coronagraph architectures detailed in the HPIC catalog, yielding an OWA of roughly 250-375 mas at λ = 1 μm.

The third critical parameter is the raw contrast ($C_{\rm raw}$), which specifies the minimum detectable planet-to-star flux ratio before post-processing \citep{Mahapatra2023,Vaughan2023}:
\begin{equation}\label{eq:contrast}
    C_{\rm raw} = \left(\frac{F_p}{F_\star}\right)_{\rm min} = \left(\frac{R_p}{r}\right)^2 \times A_g(\lambda) \times \Phi(\alpha,\lambda),    
\end{equation}
where $R_p$ is the planet radius, $r$ is the star-planet distance, $A_g$ is the geometric albedo, and $\Phi$ is the phase function. Based on requirements in \cite{Mamajek2024}, HWO targets $C_{\rm raw} = 10^{-10}$ for wavelengths from 300 nm to 2 $\mu m$.

For specific planet-star configurations, we can express the expected contrast as \citep{Kane2011}:
\begin{equation}\label{eq:flux_ratio}
\frac{F_p}{F_\star} = \left(\frac{R_p}{r}\right)^2 \times A_g(\lambda) \times \left[\frac{\sin \alpha + (\pi - \alpha)\cos \alpha}{\pi}\right], 
\end{equation}
where $\alpha$ is the phase angle between the star, planet, and observer. This expression assumes Lambertian scattering.

These specifications would place HWO's detection capabilities well beyond current ground-based facilities and position it to directly image potentially habitable worlds.

\subsection{Our Analysis of IWA and OWA} \label{subsection:iwaowa}
The IWA, OWA, and minimum contrast requirements fundamentally constrain the observable population of exoplanets. These parameters, determined by the HWO's design and performance specifications, create practical detection boundaries. While hot Jupiters exist at small orbital separations (e.g., 0.03 AU), which are typically smaller than the IWA, the primary challenge is that even at the IWA, the sensitivity is typically not as good as at the OWA. This is because no coronagraph design perfectly suppresses starlight at the inner working angle. While our analysis does not explicitly model a contrast curve that varies with angular distance from the star, this effect is implicitly accounted for by setting a minimum contrast requirement for detection. On the other hand, planets in very wide orbits, especially those viewed nearly face-on, may remain outside the OWA throughout their entire orbital period, making them undetectable. Additionally, planets in wider orbits reflect less light back to the observer due to the inverse square law, making detection increasingly challenging with orbital distance. This distance effect is especially limiting for small planets since their reflected light signal scales with their physical cross-sectional area.

\subsubsection{Methodology}
\label{subsection:methodology1}

Here, we present a geometric analysis to evaluate a wide-orbiting, giant planet's detectability across different orbital configurations. Our study encompasses 164 stars from the ExEP catalog, for which we simulated 1000 synthetic planetary orbits per star. These simulations were heavily inspired by \cite{Vaughan2023}. For each orbit, we randomly sampled the semi-major axis $a$, eccentricity $e$, inclination $i$, the argument of periastron $\omega$, and the longitude of ascending node $\Omega$ using the following distributions:

The semi-major axis was sampled log-uniformly between 5 and 30 AU
\begin{equation}\label{eq:a}
     a \sim \rm Log\mathcal{U}(5, 30), \\
\end{equation}
given the correlation uncovered from the Gemini Planet Imager Exoplanet Survey \citep[$dN/da \propto \sim a^{-2}$;][]{Nielsen2019};

Orbital eccentricities were sampled between circular and highly elliptical orbits \citep{Kipping2013}:

\begin{equation}\label{eq:ecc}
     e \sim \beta(0.867, 3.03), \\
\end{equation}

For the orbital elements, we sampled from the following distributions:
\begin{equation}\label{eq:i}
     \cos(i) \sim \mathcal{U}(-1, 1), \\
\end{equation}
where $i$ is the orbital inclination, where $i = 0$ degrees corresponds to a face-on orbit; 
\begin{equation}\label{eq:omega}
     \omega \sim \mathcal{U}(0, 2\pi), \\
\end{equation}
\begin{equation}\label{eq:Omega}
     \Omega \sim \mathcal{U}(0, 2\pi). \\
\end{equation}

We generate these orbits for different OWAs: 360 mas, 800 mas, 960 mas, 1440 mas, 1920 mas, 2280 mas, and 2520 mas. The projected angular separation for each orbit was computed considering the true star-planet separation, which varies with orbital position between periastron at $a(1-e)$ and apoastron at $a(1+e)$ for an orbit of eccentricity $e$. This approach generated 164,000 unique configurations (considering a unique inclination, eccentricity, semi-major axis, etc), allowing us to systematically evaluate planet detectability under different coronagraph working angle constraints. Using an observing wavelength of $\lambda=1\mu \rm m$ (the continuum level just to the red side of the $940\rm nm$ water feature, which is useful to look for habitable conditions), we determined whether each simulated orbit would be visible by comparing its projected separation to the telescope's OWA (see Section \ref{sec:telescope-arch} for more details). 

For a simple case, assuming small angles, the on-sky angular separation can be computed as:
\begin{equation}\label{eq:angsep}
     \theta=r_{\rm proj}/d_\star, \\
\end{equation}
where $r_{\rm proj}$ is the on-sky projected planet distance (in AU), and $d_\star$ is the planet distance from the observer (in pc). Therefore, for this simple case, the planet would be detected if the angular separation is less than the coronagraph's outer working angle: $\theta\leq \rm OWA$. In reality, the accessible angular separations would depend on the full orbital configuration, which depends on the inclination and eccentricity.

\subsubsection{The Analysis}
\begin{figure}[hbt!]
    \center
    \includegraphics[width=0.46\textwidth]{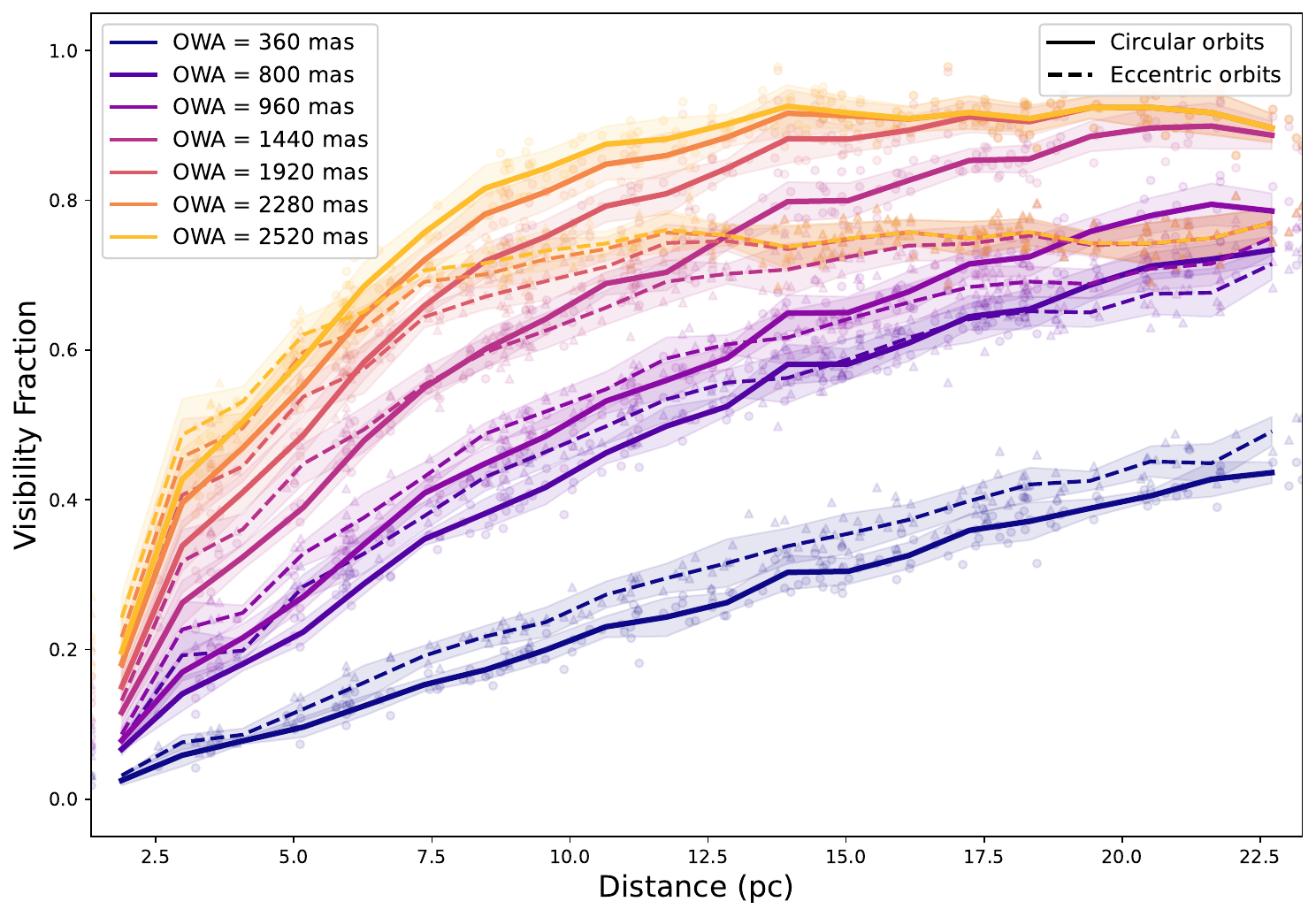}
        \caption{Planet visibility fractions as a function of stellar distance for different coronagraph configurations. The visibility fraction represents the portion of a planet's orbit during which it remains detectable between the coronagraph's IWA and OWA. Different colored lines represent various OWA values ranging from 360 to 2520 milliarcseconds (mas), with a fixed IWA of 60 mas. Solid lines show results for circular orbits ($e = 0$), while dashed lines show results for eccentric orbits ($0 < e < 1$). 
        % Note that eccentric orbits generally show lower visibility fractions compared to circular orbits at the same OWA, particularly at larger stellar distances, due to portions of their orbits falling outside the detectable range. For smaller OWAs, eccentric orbits show higher visibility because planets at periastron temporarily appear at detectable separations from their stars, while at larger OWAs, circular orbits become more visible as they maintain consistent separation throughout their orbit, whereas eccentric planets spend significant time at apastron, where they may exceed the detection limits. 
        The shaded regions represent 1$\sigma$ uncertainties derived from our orbital simulations across the ExEP stellar sample. For all cases, we assumed that the semi-major axes were uniformly distributed between 5 and 30 AU. This analysis demonstrates that an OWA of at least 800 mas is required to maintain visibility fractions above 0.6 for stars out to 20 parsecs.}
        \label{fig:iwaowa}
\end{figure}

Following the methodology outlined in Section \ref{subsection:methodology1}, Figure \ref{fig:iwaowa} shows the expected visibility fraction of planets as a function of stellar distance for different coronagraph configurations. We compare visibility fractions across different coronagraph configurations, where each colored line represents a fixed IWA of 60 mas and OWA values ranging from 360 to 2520 mas. The visibility fraction represents the probability that a planet will have at least one point in its orbit that falls within the coronagraph's detection constraints (between the inner and outer working angles), calculated across our ensemble of simulated systems with varying orbital parameters. The visibility fraction generally decreases with stellar distance for a fixed OWA, though the specific coronagraph parameters modulate the relationship. As highlighted in Figure \ref{fig:iwaowa}, larger OWAs significantly improve planet detectability, particularly for more distant stars.

Our analysis reveals that coronagraphs with small OWAs (360 mas) show rapidly declining visibility fractions closer than $\sim 5$ parsecs, making them inadequate for the majority of our target sample. At a distance of 10 parsecs, the visibility fraction drops to approximately 0.2 with a 360 mas OWA, but remains above 0.6 with a 1440 mas OWA. This difference becomes even more pronounced at 15 parsecs, where the visibility fraction falls below 0.3 for the 360 mas OWA while maintaining approximately 0.5 for an 800 mas OWA.
For coronagraphs with OWAs of 800 mas or larger, planets around stars out to 20 parsecs maintain visibility fractions above 0.7, providing adequate detection opportunities throughout most of their orbits. For the largest OWAs ($>1000$ mas) and stellar distances ($d > 10$ pc), eccentric orbits (dashed lines) have lower visibility fractions than circular orbits (solid lines). This occurs because at large OWAs, circular orbits maintain consistent separation throughout their orbit and remain within the detection limits, whereas eccentric planets spend significant time at apastron where they may exceed the coronagraph's outer working angle. On the other hand, for smaller OWAs (e.g., 360 mas), eccentric orbits show higher visibility fractions than circular orbits because planets at periastron temporarily appear at detectable separations from their stars, compensating for time spent beyond the detection limit at apastron. The effect is most pronounced for the larger OWA values, where the visibility fraction for eccentric orbits can be 10-20\% lower than for circular orbits at the same stellar distance.

We also observe that the improvement in detection completeness plateaus beyond an OWA of $\sim1400$ mas, suggesting diminishing returns for larger working angles given our target sample's distance distribution. For instance, increasing the OWA from 1440 to 2520 mas yields only a 5-10\% improvement in visibility fraction across most of our stellar distance range, while potentially increasing technical complexity and cost.
These results directly informed our recommendation for an OWA of at least 1440 mas for HWO, which would achieve detection completeness of 80-90\% for cold Jupiters across our stellar sample while balancing technical feasibility. This OWA specification ensures that the majority of planets in our target systems will be visible for a significant portion of their orbits, enabling the accurate orbital characterization demonstrated in subsequent sections of this paper.
The IWA should instead be set by the need to characterize habitable zone planets, which, according to the analysis by \cite{Morgan2024}, should fall between 20 and 65 mas. This recommendation should be able to detect most planets covered in the underlying population assumed in Section \ref{section: coldjupiters}. With these working angles, we would achieve an 80-90\% visibility fraction for cold Jupiters, which we define as giant planets with masses between 0.001-10 $M_J$ and semi-major axes of 5-30 AU. These planets have an occurrence rate of approximately 6-10\% around main sequence stars in our sample \citep{Fulton2021, Fernandes2019, Wittenmyer2020}. The cited occurrence rate studies cover a similar, though not identical, parameter space to our definition, with \cite{Fernandes2019} focusing on planets between 0.1-20 $M_J$ and separations of approximately 0.03-10 AU, while \cite{Fulton2021} examined giant planets at various orbital periods around Sun-like stars, including up to 30 AU.

\subsection{Note on Geometric Albedos}
The HabEx\footnote{\url{https://www.jpl.nasa.gov/habex/documents/}} \citep{habex} and LUVOIR\footnote{\url{https://asd.gsfc.nasa.gov/luvoir/}} \citep{Louvoir} mission proposals assumed a constant geometric albedo\footnote{the fraction of power incident on a body at a given wavelength that is scattered back out to space} of 0.2 for potentially Earth-like planets across all wavelengths studied. This value originated from \cite{Stark2014}. More recent and historical measurements suggest Earth's actual geometric albedo is higher, around 0.4. \cite{Mallama2017} found that Earth exhibits its highest albedo in ultraviolet wavelengths (0.688-0.722), with decreasing values through visible (0.512-0.434) to near-infrared (0.392-0.430) across both Johnson-Cousins and Sloan photometric systems. Jupiter, in contrast, shows peak reflectivity in visible wavelengths (0.538-0.575), with lower values in both ultraviolet (0.358-0.377) and near-infrared (0.321-0.348) regions. This aligns with earlier studies - \cite{Bakos1964} measured values of 0.41-0.42 from extensive observations in 1958-1959.

\begin{figure*}[hbt!]
    \center
    \includegraphics[width=0.9\textwidth]{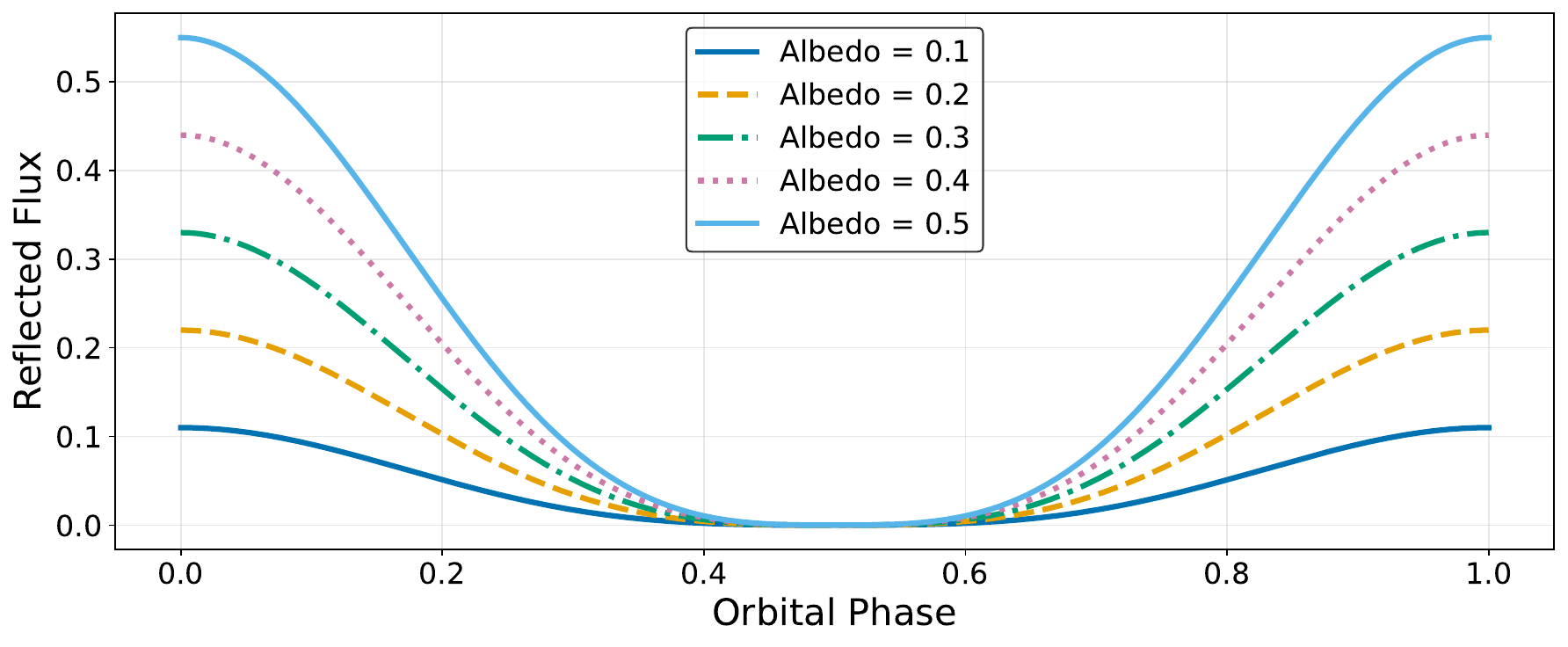}
        \caption{Phase curves showing reflected flux as a function of orbital phase for different planetary albedo values ranging from 0.1 to 0.5. The orbital phase runs from 0 to 1, where phase 0 corresponds to the inferior conjunction and phase 0.5 to the superior conjunction. All curves assume a Lambertian scattering law and identical orbital geometry. Despite the range of albedo values spanning a factor of 5, this variation is relatively minor compared to the typical dynamic range of planet-to-star contrast ratios, justifying our adoption of a single representative albedo value of 0.2 for detectability calculations.}
        \label{fig:albedos}
\end{figure*}

Here, we briefly assess the impact of planetary albedo assumptions on our detectability calculations. To do that, we modeled the reflected flux as a function of the orbital phase for different albedo values ranging from 0.1 to 0.5 (Figure \ref{fig:albedos}). It is important to examine the extrema of the curves. At transit (phase = 0.5), the observed flux drops to zero because we can only see the planet's unilluminated side. During the secondary eclipse, the total flux equals the geometric albedo of a perfectly Lambertian sphere \citep{Schwartz2015}. While the absolute reflected flux scales linearly with albedo, the relative difference between extreme albedo cases (0.1 versus 0.5) only amounts to a factor of 5 in flux. Since planet-to-star contrast ratios typically span several orders of magnitude, this relatively modest albedo-induced variation does not significantly impact our detectability predictions. We, therefore, adopt a conservative albedo value of 0.2 for our subsequent analysis, consistent with typical values for gas giants in our solar system \citep{Marley1999}. This choice allows us to maintain realistic detection predictions while acknowledging that the exact albedo values of exoplanets may vary based on their atmospheric composition and cloud coverage.

While a Lambertian sphere provides a mathematically simple model where the brightness of each surface point depends only on the combination of how directly sunlight hits it and how directly we observe it, real celestial bodies behave quite differently. Planets and moons typically exhibit complex scattering patterns that vary significantly with phase angle. For instance, atmospheric Rayleigh scattering tends to direct light forward or backward, while clouds and ocean surfaces can create strong specular reflections. Additionally, surface roughness introduces intricate scattering patterns due to small-scale surface orientations, including effects like multiple reflections between surface features and shadows cast by surface irregularities. Though there exists sophisticated research on more realistic scattering models, as demonstrated in, e.g., \cite{Heng2021}, developing a comprehensive, flexible scattering framework lies outside this paper's scope.

\begin{figure*}[hbt!]
    \center
    \includegraphics[width=0.9\textwidth]{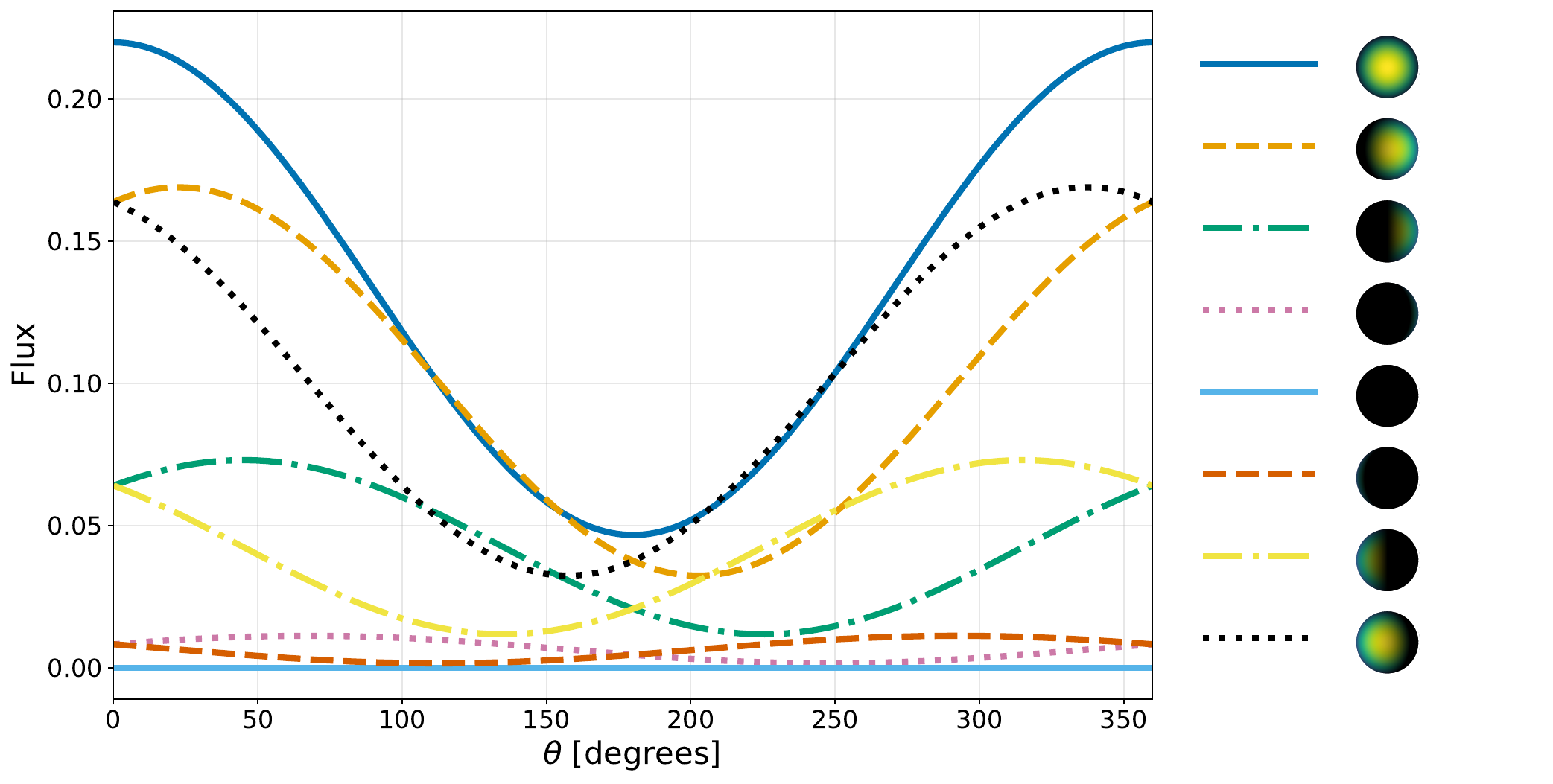}
        \caption{The phase curve of a planet over one rotation at eight different illumination phases. The images at the right show renderings of the planet's illuminated surface throughout the observation. Each curve represents a different viewing geometry, the observer-planet-star angle, with the corresponding planetary illumination phase shown by the colored circles in the legend. The top (blue) curve shows a fully illuminated planet, while the lower curves show various partial illumination scenarios down to a completely dark planet (purple). The gray curve represents the classical edge-on viewing case exhibiting the full range of phase variations.}
        \label{fig:reflected}
\end{figure*}

Unlike phase curves in thermal light, which primarily encode low-order spatial information (since the region of integration is always the full disk), phase curves in reflected light encode information at different scales depending on the phase. To illustrate how planetary orientation affects the observed reflected light, Figure \ref{fig:reflected} shows analytical light curves, derived with \texttt{starry} \citep{Luger2019}, for different viewing geometries, i.e., different phase angles corresponding to the observer-planet-star angle, of a planet with albedo 0.2. The observer is assumed to be along the ecliptic, so the illumination source is along the $x-z$ plane of a right-handed Cartesian coordinate system, with $\hat{z}$ pointing toward the observer and $\hat{x}$ pointing to the right on the sky. The axis of rotation of the modeled planet is therefore tilted clockwise away from $\hat{y}$ by $20^\circ$. Each curve represents how the reflected flux varies with orbital phase angle $\theta$ relative to the observer. The images at the right show snapshots of the planet's disk throughout the observation. When the planet is viewed edge-on, we observe the classical phase curve behavior with maximum brightness at full phase ($\theta = 0^\circ$) and minimum at new phase ($\theta = 180^\circ$). 

The shifting minima can be understood by examining how the planet's rotation interacts with varying illumination conditions. When the planet is fully illuminated (represented by the uppermost blue curve), the minimum flux occurs naturally when the darkest portion of the visible illuminated hemisphere faces the observer. However, as the illumination phase changes, represented by the progression through different colored curves, the interplay between the planet's rotation and the fraction of its illuminated surface creates distinct viewing geometries. Therefore, the minimum flux for each phase occurs at the rotation angle where the combination of illuminated fraction and planetary rotation presents the smallest visible illuminated area to the observer. This geometric effect manifests as a systematic shift in the angular position of flux minima across different illumination phases. The variation in curve amplitudes, from the highest flux in the fully illuminated phase (blue curve) to the lowest in the new/crescent phases (purple curve), directly corresponds to the fraction of the illuminated hemisphere visible to the observer at each phase. This behavior is analogous to the familiar phases of the Moon, but with the additional complexity of planetary rotation modulating the observed flux at each phase. 


\section{Our models for inferring inclination and eccentricity for Cold Jupiters} 
\label{section: coldjupiters}
Here, we present our methodology for determining the orbital inclinations and eccentricities of cold Jupiters, along with the resulting measurements and their associated uncertainties. Our approach assumes that the cold Jupiters have already been detected, and therefore, we demonstrate how to reduce uncertainties in their orbital characterization through different observational techniques.
\subsection{Methodology}
\label{subsection: methodology2}

Using the \texttt{octofitter}\footnote{\texttt{octofitter} is a Julia package for performing Bayesian inference against a wide variety of exoplanet / binary star data.} framework \citep{Thompson2023}, we ran 100 simulations comparing different star-planet systems (here, all planets are distant giant planets):
\begin{itemize}
\item RV-only model: we assumed a total of 40 RV measurements per target. The 1 m/s uncertainty was chosen as a compromise between the current precision achievable by instruments like HARPS and HIRES (around 2 m/s) and the sub-1 m/s uncertainty expected from next-generation EPRV facilities such as ESPRESSO and KPF. With higher precision, we may reduce the number of observations needed to achieve equivalent orbital constraints, optimizing telescope time.
\item Astrometry-only model: we assumed six relative astrometric measurements, each having uncertainties of 5 mas in both right ascension and declination, distributed approximately every two months over a year. The number of astrometric measurements (six) was chosen based on the results presented in Figure \ref{fig:inc-ecc-unc}, which demonstrates that the improvement in orbital parameter constraints diminishes significantly after six epochs. The astrometric uncertainty of 5 mas was chosen as a representative value for the approximate expected performance of HWO.
\item RV+Astrometry model: we are joint-modeling simulated RV data, assuming that these measurements were taken before the launch of HWO, with six relative astrometric measurements distributed approximately every two months over a year after the launch of HWO. All uncertainties stay the same. 
\end{itemize}

For our simulations, we generated synthetic planets around each star by drawing orbital parameters from carefully chosen prior distributions without consideration of any known planets in these systems. These simulated planets allow us to explore the full range of plausible orbital configurations independent of existing planetary architectures.

% For the orbital eccentricities, rather than sampling from a uniform distribution that might overrepresent highly eccentric orbits, we sampled from a beta distribution with parameters $\alpha=0.867$ and $\beta=3.03$ as suggested in \cite{Kipping2013}. However, the high eccentricities are still significant since they give implications on dynamical evolution for the planetary systems \citep{Chatterjee2008, Juric2008}. The inclination was sampled isotropically by drawing the cosine of inclination from a uniform distribution between -1 and 1, which ensures uniform sampling over the sphere. This choice of uniform sampling in $\cos{i}$ (equivalent to a $\sin{i}$ distribution) correctly represents the geometric probability distribution of inclinations, where edge-on orbits are more common than face-on configurations. This follows from the fact that there are more ways to arrange an orbit edge-on than face-on relative to our line of sight.

The prior distributions for $a$, $e$, $i$, $\omega$, and $\Omega$ follow the same distribution as specified in Section \ref{subsection:methodology1} (Equations \ref{eq:a}, \ref{eq:ecc}, \ref{eq:i}, \ref{eq:omega}, \ref{eq:Omega}). However, we note that the simulations specified in Section \ref{sec:telescope-arch} are Monte Carlo simulations where orbital element values were drawn from specified distributions. In contrast, here we fit simulated RV and relative astrometric measurements, so the distributions represent our prior knowledge that informs the inference process in the Bayesian Model.

% This range encompasses the ice line locations across the ExEP target stars out to the practical detection limits of HWO's outer working angle. While the ice line typically scales as $\sqrt{L/L_\odot}$ with stellar luminosity, varying significantly between F and M stars in our sample, we adopt a simplified uniform range for this initial analysis.

In addition to $a$, $e$, $i$, $\omega$, and $\Omega$, we also specify 
$$
\tau \sim \mathcal{U}(0,1)
$$
$$
M_p\sim\rm Log\mathcal{N}(0.001 M_J,10 M_J)
$$

where $\tau$ is the dimensionless phase parameter indicating the fraction of orbital period elapsed since periastron passage, and $M_p$ is the planet mass in Jupiter masses $M_J$. The orbital period is derived from the semi-major axis and stellar mass using Kepler’s third law.

To sample from these prior distributions and explore the posterior parameter space, we employed the Hamiltonian Monte Carlo (HMC) algorithm with No-U-Turn Sampling \citep[NUTS;][]{Hoffman2011}. HMC-NUTS was chosen for its efficient exploration of the parameter space and superior handling of the high-dimensional orbital geometry \citep{betancourt2017, salvatier2015}.

The simulations were implemented in \texttt{julia} using the \texttt{octofitter} package's built-in HMC-NUTS sampler \citep{xuAdvancedHMCJlRobust2020}.
For each simulation, we drew 1,000 posterior samples, following a warm-up phase of 1,000 iterations. We initialized the chains using Multi-pathfinder \citep{Zhang2022}. This approach repeatedly draws random points from the priors, optimizes them while storing the optimization traces, and uses these traces to construct a variational approximation of the posterior. The different optimization runs are then combined as a mixture of Gaussians, from which we draw to initialize the sampler. This initialization strategy provides robust starting points that help prevent the sampler from getting stuck in local optima in multimodal posteriors.

\subsection{Results}

\begin{figure}[hbt!]
    \center
    \includegraphics[width=0.46\textwidth]{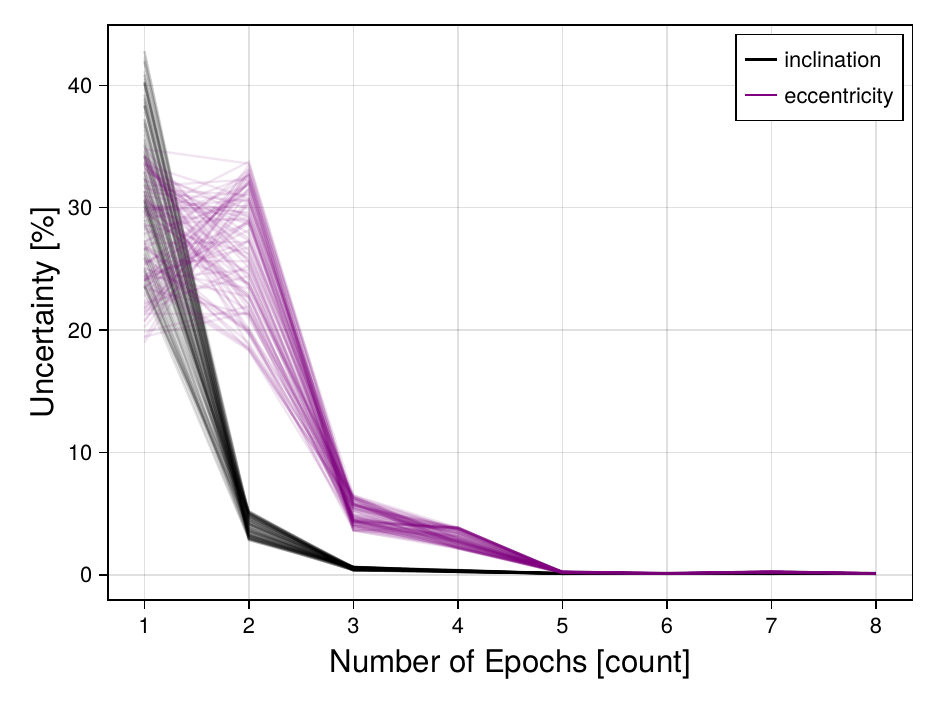}
        \caption{Reduction in the uncertainty in \% of orbital parameters as a function of the number of astrometric observations (epochs) with an uncertainty of $1^\circ$ for 100 simulated planets. The black lines show the uncertainty in orbital inclination, while the purple lines show the uncertainty in eccentricity. Both parameters show rapid improvement with initial observations, with uncertainties dropping significantly after five epochs. Additional observations beyond six epochs yield diminishing returns, suggesting this as an optimal number of visits for orbital characterization with HWO for wide-separation planets.}
        \label{fig:inc-ecc-unc}
\end{figure}

Our orbital parameter recovery analysis demonstrates that multiple astrometric observations are essential for constraining planetary orbital geometry. Figure \ref{fig:inc-ecc-unc} shows how measurement uncertainties in both eccentricity and inclination decrease as the number of astrometric observations increases. While RV observations alone provide no or limited constraints on these geometric parameters, we find that approximately eight well-timed astrometric measurements with HWO can reduce these uncertainties to $\sim \pm 3-6^\circ$. Therefore, to characterize planetary orbits to better than $\sim\pm3-6^\circ$ precision (in inclination), we should aim to obtain at least 6-8 astrometric measurements, with timing optimized to sample different orbital phases. 

\begin{figure*}[hbt!]
    \center
    \includegraphics[width=0.9\textwidth]{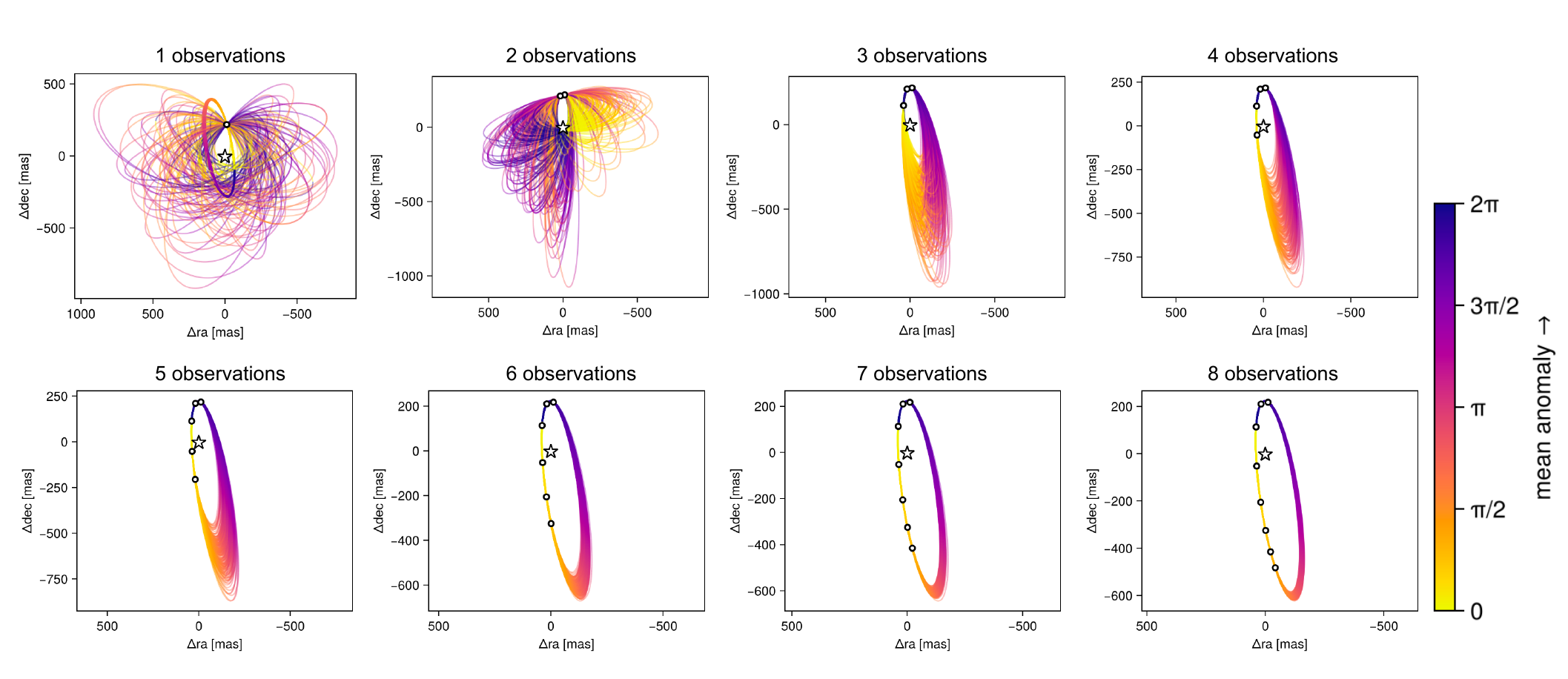}
        \caption{Evolution of orbital solution constraints with an increasing number of simulated astrometric observations, shown in relative right ascension ($\Delta \rm ra$) and declination ($\Delta \rm dec$) coordinates in milliarcseconds. Each panel shows the ensemble of possible orbits (colored lines) consistent with the simulated measurements (black circles) for different numbers of observations from 1 to 8. The star symbol indicates the position of the star. The color scale represents the mean anomaly from 0 to $2\pi$. The progressive reduction in the spread of possible orbits demonstrates how additional observations improve orbital constraints, with significant convergence achieved by six observations.}
        \label{fig:8-orbits}
\end{figure*}

Figure \ref{fig:8-orbits} illustrates how our ability to constrain planetary orbits improves with an increasing number of astrometric observations through a series of simulated orbital fits. Each panel represents the results of our orbital fitting procedure using different numbers of simulated observations, from one to eight measurements, for a planet at 5 AU and with an eccentricity of 0.6 orbiting a star at 20 pc. We chose this specific configuration as a representative case to establish baseline observational requirements, although we acknowledge it doesn't represent the whole parameter space of distant giant planets. The star symbol represents the simulated star's position, while white circles indicate the locations of the simulated measurements. The color of each orbital solution represents the mean anomaly of the planet. The orbital solutions are highly degenerate with just one or two observations, as shown by the scattered distribution of possible orbits in the top panels. As more observations are added, the orbital solutions begin to converge toward the true orbit. The dramatic improvement between one and six observations demonstrates the importance of multiple epoch measurements for accurate orbital characterization. After six observations, the additional measurements provide only incremental improvements in orbital constraints, suggesting this as an optimal number of visits for HWO's observing strategy. If the planet has some precursor RV measurements, then even six astrometric measurements could be enough for the level of precision of $<10\%$. For planets with longer orbital periods (approaching 30 AU), more extensive temporal coverage would be required, as discussed further in this Section and Section \ref{sec:discussion}.

\begin{figure*}[hbt!]
    \center
    \includegraphics[width=0.9\textwidth]{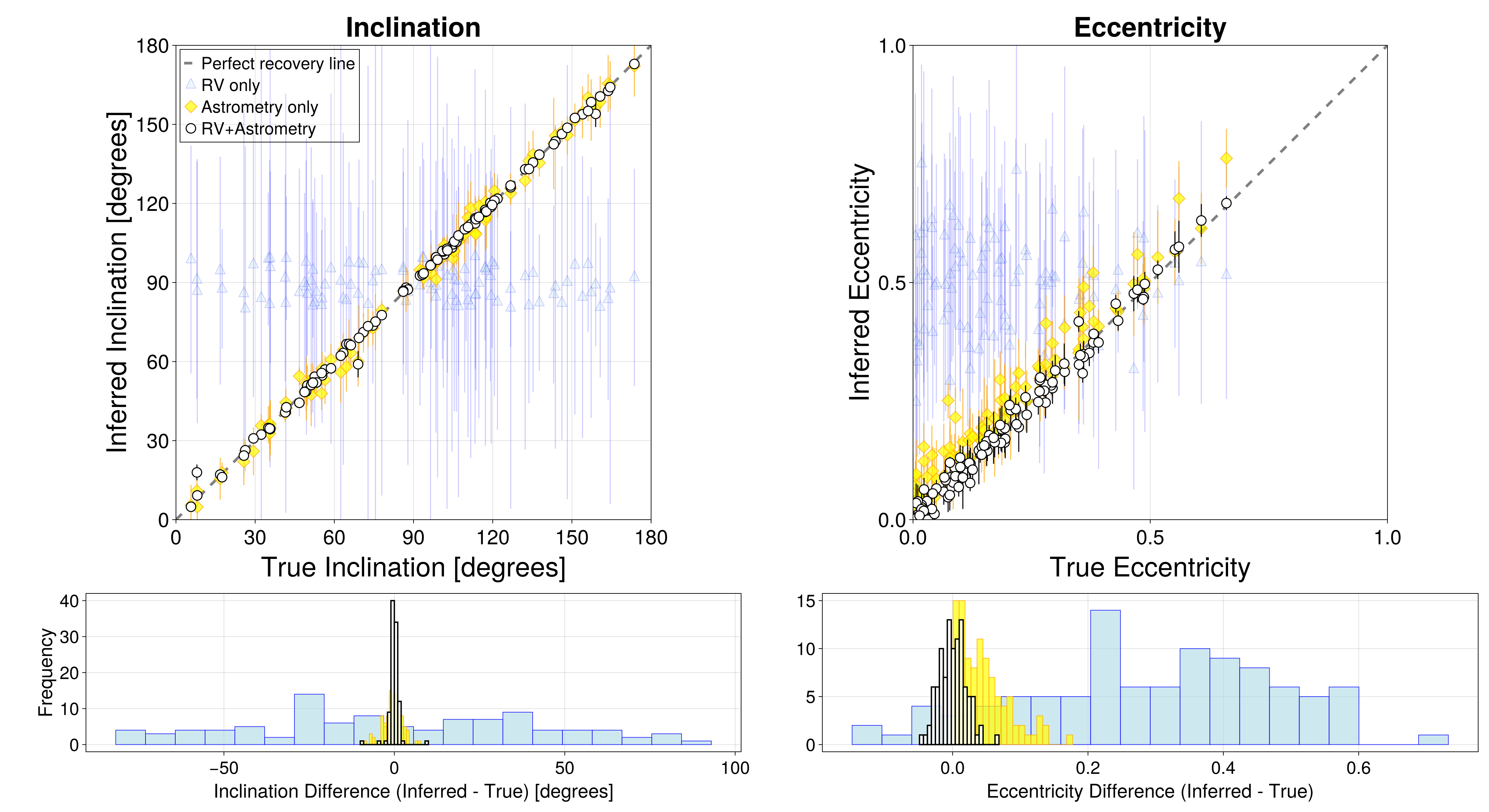}
        \caption{The accuracy of exoplanet orbital parameter determination using radial velocity (RV) measurements combined with astrometry. The top panels show the correlation between inferred and true values for orbital inclination (left) and eccentricity (right), with blue triangles representing RV-only measurements, yellow diamonds representing Astrometry-only measurements, open circles showing combined RV+Astrometry measurements, and the dashed line indicating perfect recovery. The error bars show $1\sigma$ uncertainty. The bottom panels display histograms of the measurement errors (inferred minus true values) for inclination (left) and eccentricity (right), with blue bars showing the RV-only results, yellow bars showing Astrometry-only results, and white bars showing combined RV+astrometry results. The combined approach significantly improves parameter recovery, where RV alone shows large systematic uncertainties, while both methods achieve relatively good precision.}
        \label{fig:inc-ecc-comparison}
\end{figure*}

RV measurements alone cannot fully constrain a planetary orbit's three-dimensional geometry due to inherent degeneracies between inclination and planetary mass, which is why precise direct imaging astrometry is essential for breaking these degeneracies and determining the true orbital parameters. However, when we have precursor RV measurements available, we can significantly improve our astrometric solutions by leveraging the complementary strengths of both techniques, allowing for better constraints on orbital parameters even with relatively sparse astrometric sampling. We show in Figure \ref{fig:inc-ecc-comparison} that the substantial improvement in orbital parameter determination is achieved by combining precursor radial velocity measurements with astrometric observations. The upper panels present scatter plots comparing inferred versus true values for orbital inclination (left) and eccentricity (right). Blue triangles represent results from radial velocity data alone, yellow diamonds show the results for Astrometry-only measurements,  while open circles show the combined RV+astrometry approach. The dashed diagonal line indicates perfect parameter recovery. For orbital inclination, RV measurements alone exhibit significant systematic biases and large uncertainties. The RV-only method shows substantial scatter around the perfect recovery line, with many measurements deviating significantly from true values. In contrast, both the Astrometry-only and the combined RV+astrometry approach closely follow the correlation, demonstrating markedly improved accuracy across the full range of inclination angles. Eccentricity determination shows better performance for both Astrometry and RV+Astrometry methods, though the combined approach still provides superior results. The lower panels quantify these improvements through histograms of measurement residuals (inferred minus true values). For inclination, the RV-only method (blue) produces a broad, asymmetric distribution with a significant systematic offset, while the combined approach (white) yields a much narrower, more symmetric distribution centered near zero. The eccentricity residuals show again that both Astrometry-only and RV+astrometry methods achieve reasonable precision, but the combined RV+astrometry approach produces a tighter distribution with reduced systematic bias. These results highlight the critical importance of incorporating astrometric measurements for robust exoplanet orbital characterization, particularly for determining orbital inclination where RVs alone provide limited constraints.

\begin{figure*}[hbt!]
    \center
    \includegraphics[width=0.9\textwidth]{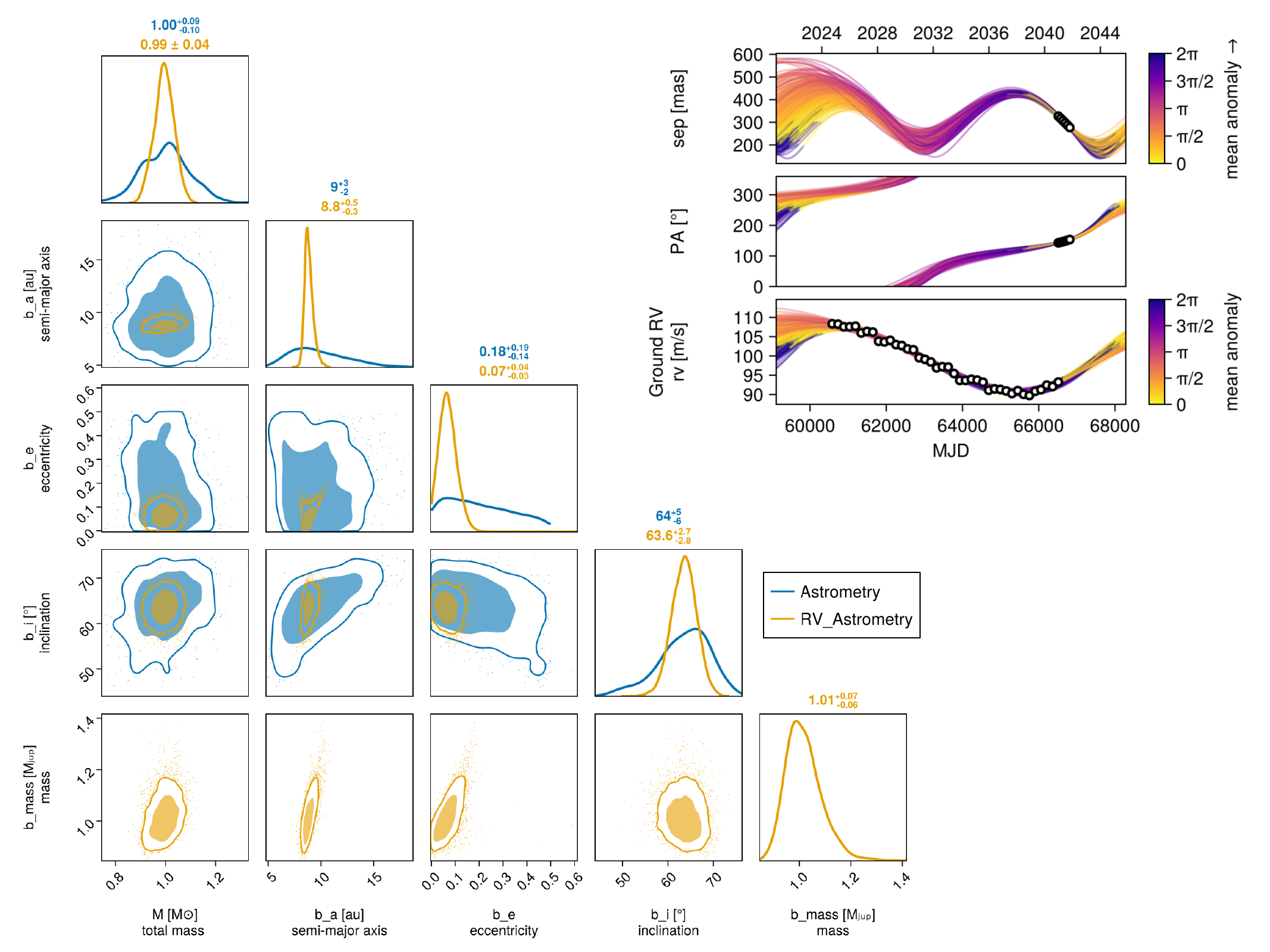}
        \caption{Corner plot showing posterior distributions of orbital parameters from two different models: Astrometry-only (blue) and combined RV+astrometry (orange) for a simulated planetary system. The star is at 20 pc, and the astrometric measurements span two months between epochs. The top right panels show the predicted astrometric separation (in mas), position angle (in degrees), and RV measurements (in m/s) over time. The corner plot displays joint and marginal posterior distributions for the total mass (\texttt{M}), planetary semi-major axis (\texttt{b\_a}), eccentricity (\texttt{b\_e}), and inclination (\texttt{b\_i}). The true system parameters correspond to a nearly circular orbit ($e = 0.07$) at inclination ($i = 64^\circ$). The addition of precursor RVs to the astrometric data significantly improves parameter constraints, particularly for eccentricity and semi-major axis.}
        \label{fig:astrometry_rvastrometry}
\end{figure*}

To see this more in depth, Figure \ref{fig:astrometry_rvastrometry} demonstrates the significant improvement achieved by combining RV and astrometric measurements compared to astrometry alone for orbital parameter determination. The corner plot displays joint posterior distributions for key orbital parameters (total mass, semi-major axis, eccentricity, and inclination), with blue representing astrometry-only constraints and orange indicating the combined RV+astrometry solution. The top panels display the predicted angular separation, position angle evolution, and ground-based RV measurements for approximately 15-20 years (assuming the RVs are taken before the HWO's operation). The combined solution (orange) provides considerably tighter constraints on all parameters, with the eccentricity being refined from 0.18 to $0.07\pm0.04$, and the inclination narrowing from a broad distribution around $64^\circ$ with astrometry alone to a more precise $63.6^\circ\pm2.7^\circ$ when including RV data. Similarly, the semi-major axis uncertainties improve from an AU to a sub-AU level, demonstrating how precursor RV measurements significantly enhance the precision of orbital solutions even with limited astrometric sampling.

\begin{figure*}[hbt!]
    \center
    \includegraphics[width=0.9\textwidth]{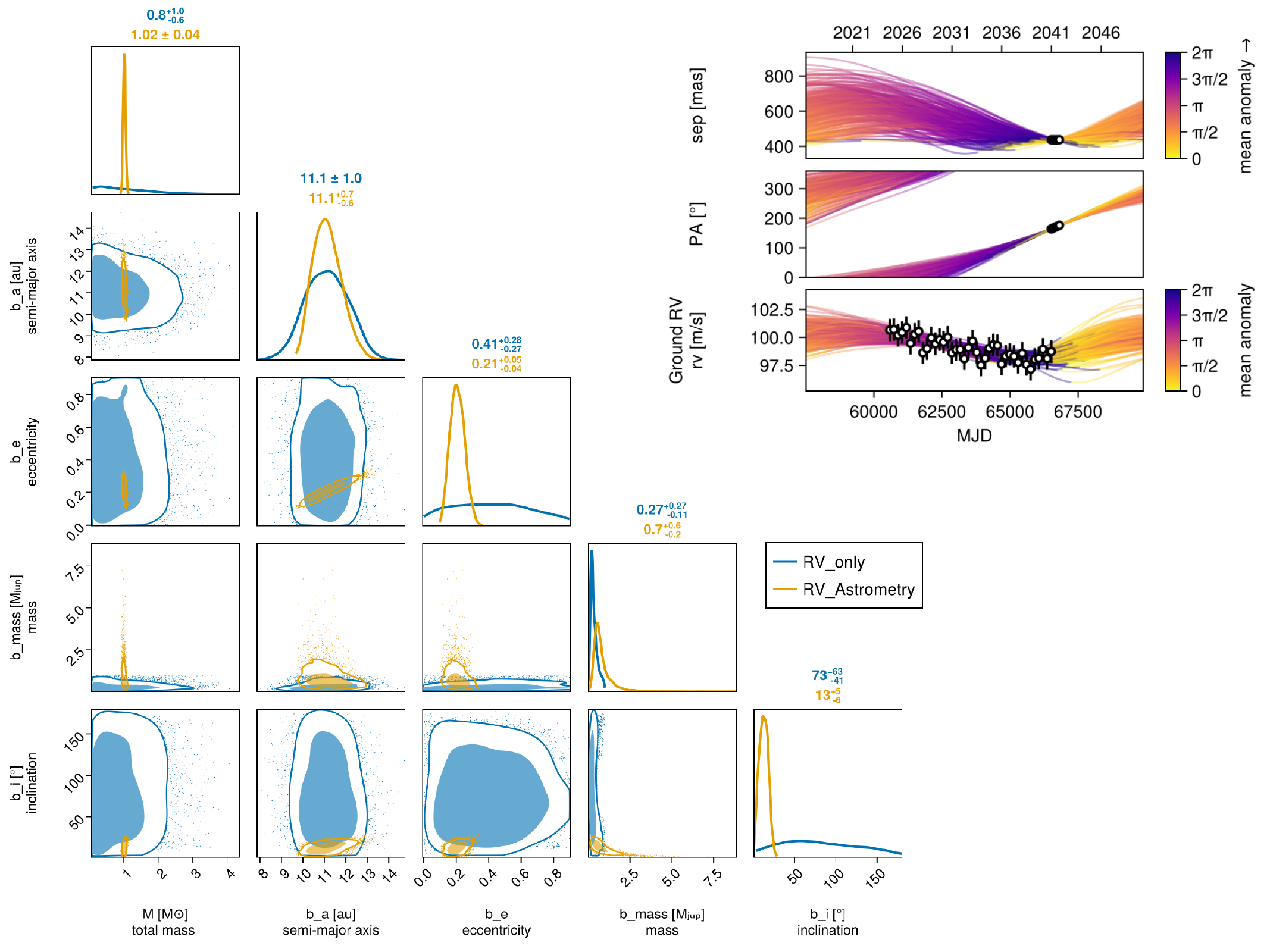}
        \caption{Corner plot showing posterior distributions of orbital parameters from two different models: RV-only (blue) and combined RV+astrometry (orange) for a simulated planetary system. The star is at 20 pc, and the astrometric measurements span two months between epochs. The top right panels show the predicted astrometric separation (in mas), position angle (in degrees), and RV measurements (in m/s) over time. The corner plot displays joint and marginal posterior distributions for the total mass (\texttt{M}), planetary semi-major axis (\texttt{b\_a}), eccentricity (\texttt{b\_e}), inclination (\texttt{b\_i}), and planetary mass (\texttt{b\_mass}). The true system parameters correspond to a quite eccentric orbit ($e = 0.2$) at low inclination ($i = 10^\circ$). The addition of astrometric measurements to the RV data significantly improves parameter constraints, particularly for inclination and eccentricity.}
        \label{fig:rv_rvastrometry}
\end{figure*}

At the same time, Figure \ref{fig:rv_rvastrometry} demonstrates the comparison between having only RV observations and RV+astrometry observations for a different system in our sample. This time it is an eccentric planet ($e = 0.2$) with an inclination of $i = 10^\circ$ (these parameters were chosen randomly, but we wanted to illustrate a completely different case from what is shown in Figure \ref{fig:astrometry_rvastrometry}). The figure compares two orbital solutions: one using only simulated RV measurements with uncertainties of 1 $\rm m/s$ (shown in blue) and another combining RV with astrometric observations (shown in orange). The top right panels display the predicted astrometric signature (angular separation versus time), the position angle evolution, and the RV measurements spanning approximately 20 years. The corner plot below shows the joint posterior distributions for fundamental orbital parameters, including the total mass, planetary semi-major axis, eccentricity, and inclination. The RV-only model exhibits significant degeneracies between parameters, particularly in constraining the inclination and eccentricity. However, adding six astrometric measurements dramatically improves our constraints, collapsing the posterior distributions to narrow ranges around the true values. This is particularly evident in the inclination determination, which sharpens from a broad distribution in the RV-only case to a precise measurement of $\sim13^\circ$ in the combined solution, subsequently providing a definitive planetary mass measurement rather than just a minimum mass. The high eccentricity of this system is also much better constrained in the combined solution \citep[see, e.g.,][]{Reffert2011}, demonstrating how multi-technique observations are crucial for characterizing complex orbital architectures.

\begin{figure*}[hbt!]
    \center
    \includegraphics[width=0.9\textwidth]{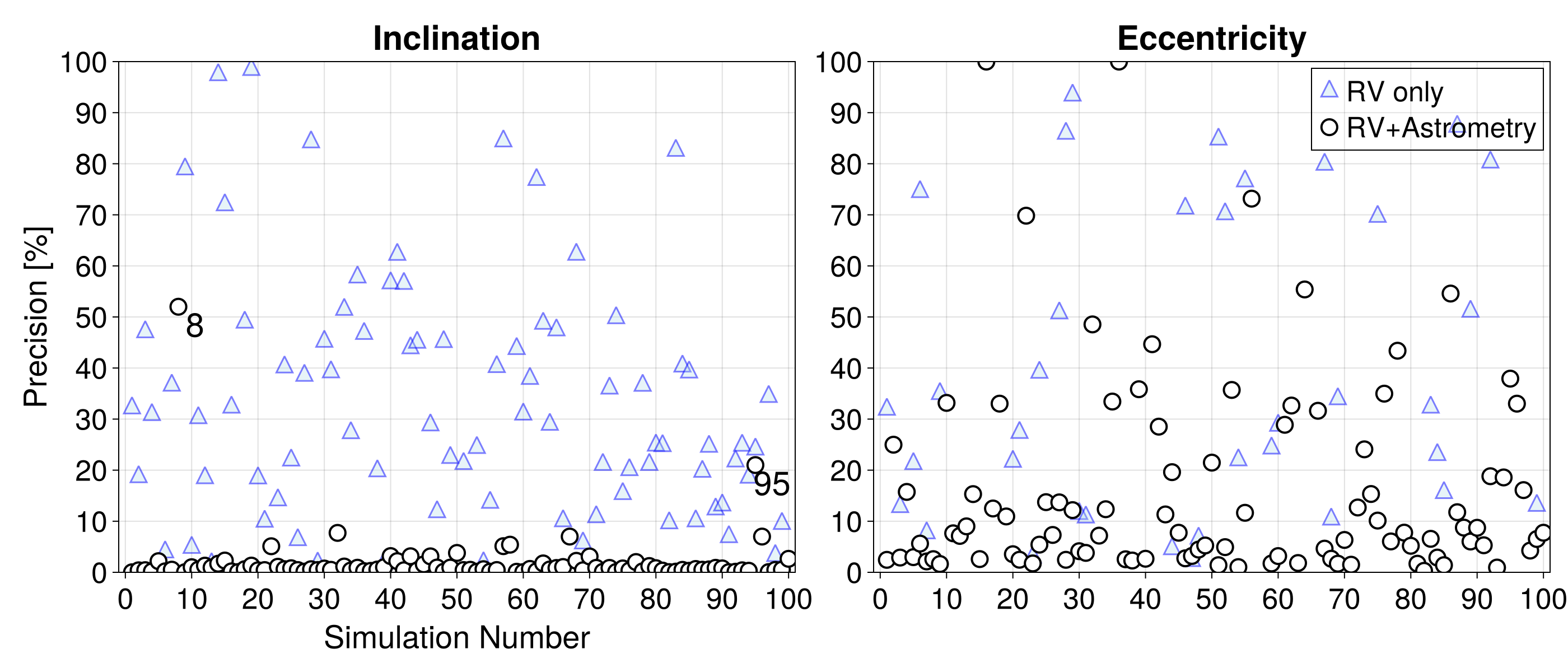}
        \caption{Parameter recovery precision across 100 independent orbital simulations combining RV and astrometric measurements after six epochs of astrometric observations covering around a year (one epoch every two months). The left panel shows relative precision in recovered inclination values, while the right panel shows uncertainties in eccentricity. Each point represents the percentage uncertainty for a single simulation. The varying levels of precision reveal a key limitation in orbital characterization: planets with periods significantly longer than our observational baseline show markedly higher uncertainties. We highlight several cases (numbered 8 and 95) with uncertainties reaching up to 50\% for inclination and 60\% for eccentricity. These outliers share a common feature - their orbital periods (128 and 133 years, respectively) are much longer than our observational baseline, resulting in coverage of only a small fraction of their complete orbits. This temporal sampling limitation, rather than specific inclination values or other orbital elements, proves to be the primary factor determining parameter recovery precision.}
        \label{fig:inc-ecc-precision}
\end{figure*}

Further, we measure the relative precision (or relative error) in \% in eccentricity and inclination determinations using the methodology described in Section \ref{subsection: methodology2}.
Figure \ref{fig:inc-ecc-precision} shows the percentage uncertainties in the recovered inclination and eccentricity values across 100 simulations for different systems. Most orbits show good precision with relative uncertainties below 10\%, indicating robust orbital solutions. However, we identified several cases (numbered in the inclination panel) where the precision deteriorates significantly, with relative uncertainties reaching up to $\sim 50$\%. These outlier cases warrant a closer examination of their orbital characteristics to understand the factors limiting precise parameter determination. 

% Our detailed analysis revealed a fundamental limitation: the ratio between the orbital period and observational baseline. The outlier cases have orbital periods exceeding 100 years, whereas our simulated observations span approximately one year. 

%

\begin{figure*}[hbt!]
    \center 
    \includegraphics[width=0.9\textwidth]{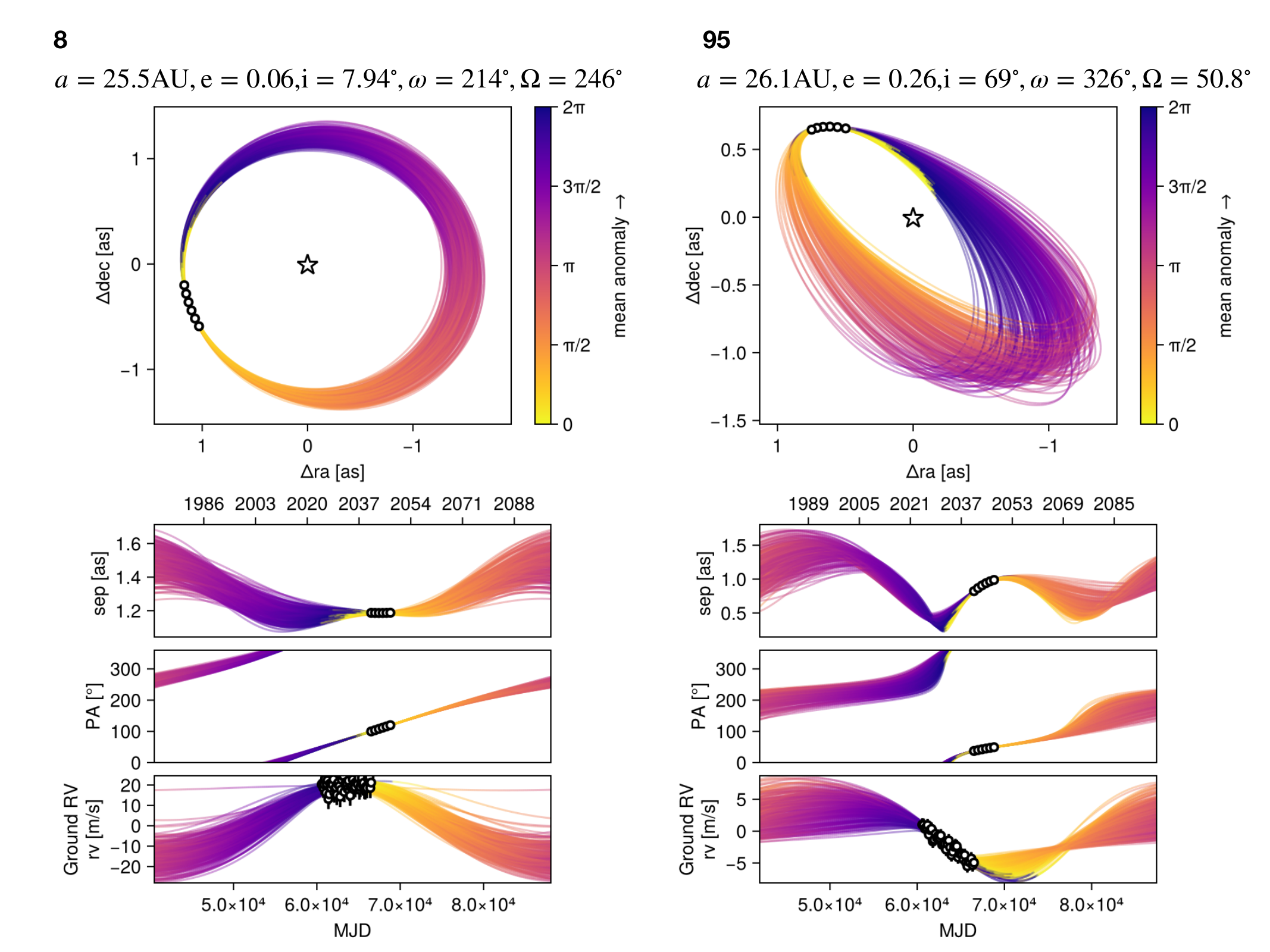}
        \caption{Orbital fits for the two cases showing poor precision in their orbital parameters (simulations 8 and 95). Each panel displays the projected orbit in the plane of the sky, where $\Delta \rm ra$ and $\Delta \rm dec$ are given in milliarcseconds (mas). The white circles represent the actual observations, while the star symbol indicates the position of the central body. The color gradient along the orbit traces the mean anomaly from 0 (yellow) to $2\pi$ (dark blue), illustrating the object's motion along its orbit. The orbital parameters are shown above each panel: semi-major axis ($a$) in AU, eccentricity ($e$), inclination ($i$) in degrees, the argument of periapsis ($\omega$) in degrees, and longitude of ascending node ($\Omega$) in degrees. These problematic cases all share extremely long orbital periods with 128 years in simulation 8, and approximately 150 years in simulation 95, resulting in our observations covering only 1-2\% of a complete orbit. This limited orbital coverage is the primary factor limiting precise parameter determination.}
\label{fig:bad-precisions}
\end{figure*}

Figure \ref{fig:bad-precisions} provides a closer look at these challenging cases. Each panel shows the projected orbit in the plane of the sky, with white circles representing the actual observations and the star symbol marking the central body. The color gradient along each orbit indicates the mean anomaly from 0 to $2\pi$, effectively showing how the object moves along its orbit with time. The orbital parameters for each case are listed above their respective panels, including the semi-major axis ($a$), eccentricity ($e$), inclination ($i$), the argument of periapsis ($\omega$), and longitude of ascending node ($\Omega$). These problematic cases represent orbits with extremely long periods relative to our observational timeline. For instance, simulation 8 has an orbital period of approximately 130 years, meaning our one-year observational campaign samples less than 1\% of the complete orbit. This limited orbital coverage fundamentally restricts our ability to constrain the complete orbital parameters, regardless of the orbit's orientation. The colored orbital pathways show the full range of possible orbits consistent with our limited observations, highlighting the substantial uncertainty that remains when such a small orbital arc is sampled. 

Our detailed examination focused on simulations that exhibited uncertainties exceeding 10\% in their orbital parameters (simulations 8 and 95). However, our initial analysis focused primarily on \textit{relative} uncertainties in orbital parameters. When examining \textit{absolute} uncertainties, particularly in orbital inclination, we identified several notable ``outlier'' cases where the uncertainty in inclination ($\Delta i$) exceeds 10 degrees. Simulation 95 overlaps with the outliers previously identified in our relative uncertainty analysis (Figure \ref{fig:inc-ecc-precision}), suggesting systematic challenges in parameter determination for certain orbital configurations. We, therefore, concentrated our detailed analysis on some specific cases (simulations 8 and 95), as they appear to represent particularly challenging scenarios for combined RV and astrometric orbit determination. 

\begin{figure*}[hbt!]
    \center
    \includegraphics[width=0.9\textwidth]{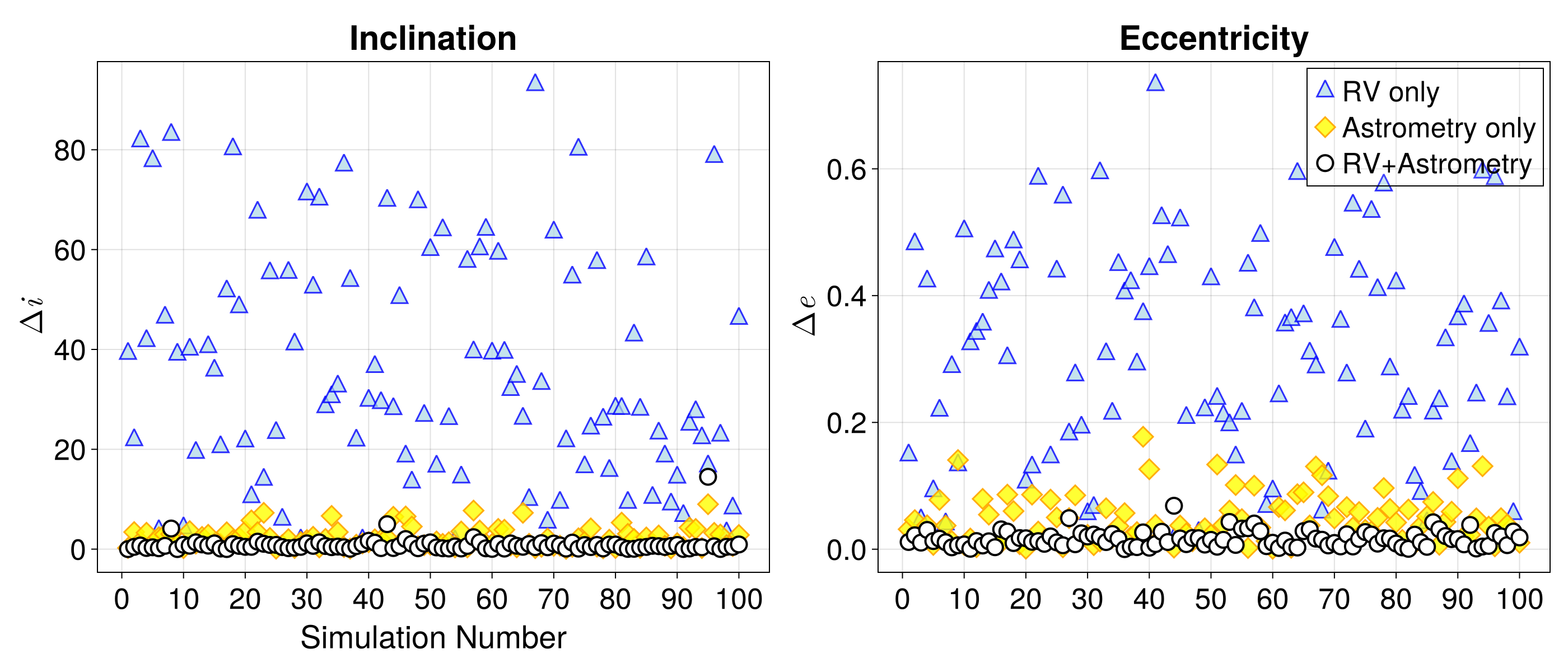}
        \caption{Comparison of eccentricity absolute uncertainties ($\Delta e$) and inclination absolute uncertainties ($\Delta i$) between RV-only (blue triangles), Astrometry-only (yellow diamonds), and combined RV+astrometry (white circles) solutions across 100 simulated planetary systems. While the RV+astrometry solutions show higher uncertainties in some cases, particularly for nearly circular orbits, the absolute uncertainty for eccentricity remains below 0.1.}
        \label{fig:ecc-abs}
\end{figure*}

To address the high relative precision errors for eccentricity, we argue that relative precision is not a sufficient metric for nearly circular orbits, and one should instead examine absolute uncertainties to properly assess the model's performance and the physical significance of the results. For a planetary system with a true eccentricity of 0.05, an absolute measurement error of 0.03 represents relatively good absolute precision but corresponds to a 60\% relative error. However, a system with true eccentricity of 0.8 could have the same 0.03 absolute error, yielding only a 3\% relative uncertainty. This effect is particularly pronounced because many samples in our simulations have low eccentricities due to the chosen prior distribution (Equation \ref{eq:ecc}). Figure \ref{fig:ecc-abs} shows that Astrometry-only measurements achieve more consistent precision, with most eccentricity errors falling below 0.2. The combined approach again provides the best results, with eccentricity errors predominantly below 0.1.

In our initial analysis, the six astrometric measurements were scheduled at two-month intervals within a single year. However, re-analysis of simulations 8 and 95, with observations randomly distributed across the full five-year mission timeline, showed dramatic improvements in precision. 

\begin{figure*}[hbt!]
    \center
    \includegraphics[width=0.9\textwidth]{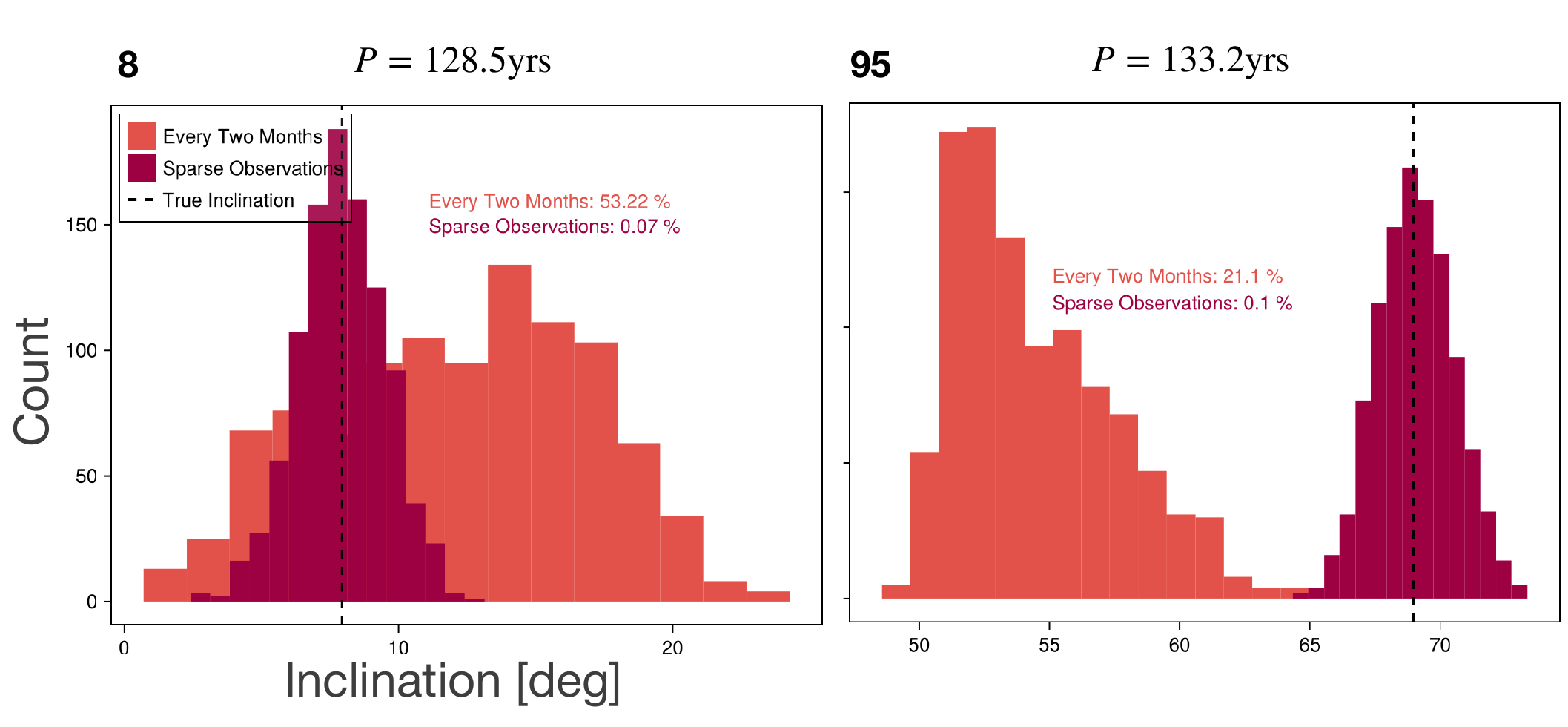}
        \caption{Comparison of orbital inclination constraints for four planetary systems using different observation timing strategies. Histograms show the posterior distributions of inclination measurements for observations scheduled every two months within one year (light red) and observations randomly distributed across the five-year mission timeline (dark red). The true inclination for each system is shown by the vertical dashed line. Each strategy's orbital period ($P$) and measurement uncertainties are indicated. All four cases demonstrate significant improvement in precision when using sparse sampling over the extended baseline, with uncertainty reductions of 1-3 orders of magnitude. This improvement is achieved despite still not covering the full orbit in longer-period systems, but covering a good fraction of it with this sampling, highlighting the importance of extended temporal baselines for orbital characterization.}
        \label{fig:sparse}
\end{figure*}

Figure \ref{fig:sparse} demonstrates the impact of observation timing strategies on orbital inclination measurements for two cases with large errors in Figures \ref{fig:inc-ecc-precision} and \ref{fig:bad-precisions}(simulations 8 and 95) with varying orbital periods. We compare two observing strategies for each simulation: measurements taken every two months within a single year (light red) and observations randomly distributed across the five-year mission timeline (dark red). The true inclination value for each system is indicated by the vertical dashed line. The percentage values shown represent the measurement uncertainties achieved with each strategy. Each system's orbital periods ($P$) are $128.5$ years and $133.2$ years, respectively, illustrating how temporal sampling affects our ability to constrain orbital parameters across different timescales.
This comparison reveals that sparse sampling over the longer baseline consistently produces more accurate inclination measurements, with uncertainties decreasing by orders of magnitude compared to the clustered observations. In simulation 8, the uncertainty drops from $53.22$\% to just $0.07$\%. In simulation 95, spreading the observations across five years significantly improves the precision, reducing the uncertainty from $21.1$\% to $0.1$\%.

Therefore, with further investigation, we found that these higher uncertainties primarily stem from insufficient orbital coverage during the observation window, directly linked to the planets' orbital periods. This result suggests that while HWO's proposed observation strategy of 6-8 astrometric measurements works well for planets with orbital periods under 100 years, the temporal distribution of these observations is crucial. Rather than clustering observations within a short timeframe, spreading measurements across the full mission timeline significantly improves orbital constraints for most targets, especially for high orbital periods. 

In addition to the constraint on the coverage of the orbital fraction, we want to draw attention to other orbital configurations that cause various degeneracies and are worth paying attention to while determining orbital parameters from RV+astrometry. These degeneracies arise when different orbital configurations produce similar observable signatures, making it challenging to determine the true orbital parameters uniquely. In nearly face-on orbits (low inclination), the projected motion becomes almost circular regardless of the true orbital shape. This geometric effect creates an inherent degeneracy between eccentricity and the argument of periastron, as the apparent orbital motion loses the distinctive signatures of ellipticity. The challenge becomes particularly acute when combined with moderate to high eccentricities, as the non-uniform orbital velocity adds another layer of complexity to the parameter determination.
The orientation of the orbit in three-dimensional space, defined by the argument of periastron ($\omega$) and the longitude of ascending node ($\Omega$), can either reduce or enhance these degeneracies. When these angles align such that the periastron passage occurs near the line of nodes (similar values of $\omega$ and $\Omega$), the maximum orbital velocity coincides with the crossing of the sky plane. This configuration can make it particularly difficult to constrain orbital parameters, as it reduces the distinctive signatures in both radial velocity and astrometric measurements that help break parameter degeneracies.
While the combination of radial velocity and astrometric measurements generally helps to break many of these degeneracies by providing complementary information about the orbital motion, certain geometric configurations remain challenging. The most problematic cases typically arise from combinations of:
\begin{itemize}
    \item low inclination angles that reduce the projected ellipticity,
    \item moderate to high eccentricities that create non-uniform orbital motion,
    \item and particular alignments of $\omega$ and $\Omega$ that limit our ability to leverage the complementary nature of radial velocity and astrometric measurements.
\end{itemize}

\section{Our Models for inferring orbits of HZ planets}
\label{section:hzplanets}
To assess HWO's capability to characterize HZ planets, we conducted simulations of single-planet systems around the stars from NASA's Exoplanet Exploration Program Mission List \citep{Mamajek2024}. 
\subsection{Methodology}
\label{subsection: HZmethod}

While we maintained most orbital parameter distributions as described in Section \ref{subsection: methodology2} \textit{for the cold Jupiter simulations}, we modified the semi-major axis distribution to specifically target potentially habitable planets. We adopted a log-uniform distribution between $a_1 = 0.95\, \rm AU$ and $a_2 = 5\, \rm AU$, with values scaled by the square root of stellar luminosity to account for the varying location of habitable zones across different stellar types.

We defined planets as ``fully within the habitable zone" when both their perihelion and aphelion fell within the optimistic habitable zone boundaries. These boundaries were calculated as $\mathrm{HZ}_{\rm inner} = \sqrt{L/1.78}$ AU for the inner edge and $\mathrm{HZ}_{\rm outer} = \sqrt{L/0.32}$ AU for the outer edge, where $L$ is in units of the solar luminosity $L_{\odot}$ \citep{kopparapu2013a,kane2016c}.

Our observing strategy, designed to simulate potential HWO observations, consisted of eight observations scheduled at two-month intervals. This specific cadence of eight observations at two-month intervals applies only to the simulated HWO direct imaging observations. For detection, we required that the planet-star contrast, computed in reflected light at a wavelength of 1 $\mu \rm m$, exceed $10^{-10}$ and that the projected separation fall within working angles of 62 for IWA to 2000 mas for OWA (see Section \ref{subsection:iwaowa}). We employed a two-stage orbital fitting approach to analyze these observations. During early epochs, when orbital coverage was limited, we utilized Orbits for the Impatient \citep[OFTI,][]{Blunt2017}, a Bayesian rejection-sampling algorithm specifically designed for undersampled cases. As additional epochs provided better orbital constraints, we transitioned to a Metropolis-Hastings Markov Chain Monte Carlo (MCMC) algorithm \citep{Metropolis1953, hastings1970, Nielsen2014}, which we ran to convergence to generate 100,000 orbital parameter posteriors.

To balance computational efficiency with statistical robustness, we randomly selected 500 posteriors from this distribution for our analysis. After each observational epoch, we analyzed these fitted orbits to compute the fraction that remained fully within the habitable zone. This fraction served as our confidence metric for confirming a planet's habitable zone status. For instance, we considered a planet to be confirmed within the habitable zone at 95\% confidence if 475 or more of the fitted orbits lay within the habitable zone boundaries.

\subsection{Results}

A planet is considered to reside in the HZ of its host star if liquid water could exist on its surface, assuming an Earth-like planet with a $\mathrm{CO_2/H_2O/N_2}$-rich atmosphere \citep{Kasting1993,kane2012a,kopparapu2013a,kopparapu2014,hill2023}. However, simply detecting a planet at a projected separation consistent with the HZ is only the first step in assessing habitability potential. Planets with non-zero eccentricities may spend only part of their orbits in the HZ, experiencing temperature swings that could preclude stable surface liquid water \citep{Williams2002,kane2012e}. Thus, confirming HZ residency requires precise constraints on the planet’s semimajor axis and eccentricity. Here, we assess the observational requirements for confirming HZ status by quantifying how many epochs are needed to characterize a planet’s orbit.

\begin{turnpage}               
  \begin{figure*}[p]            
    \centering
    \includegraphics[width=1.05\textheight, height=\dimexpr\textwidth-10\baselineskip\relax]{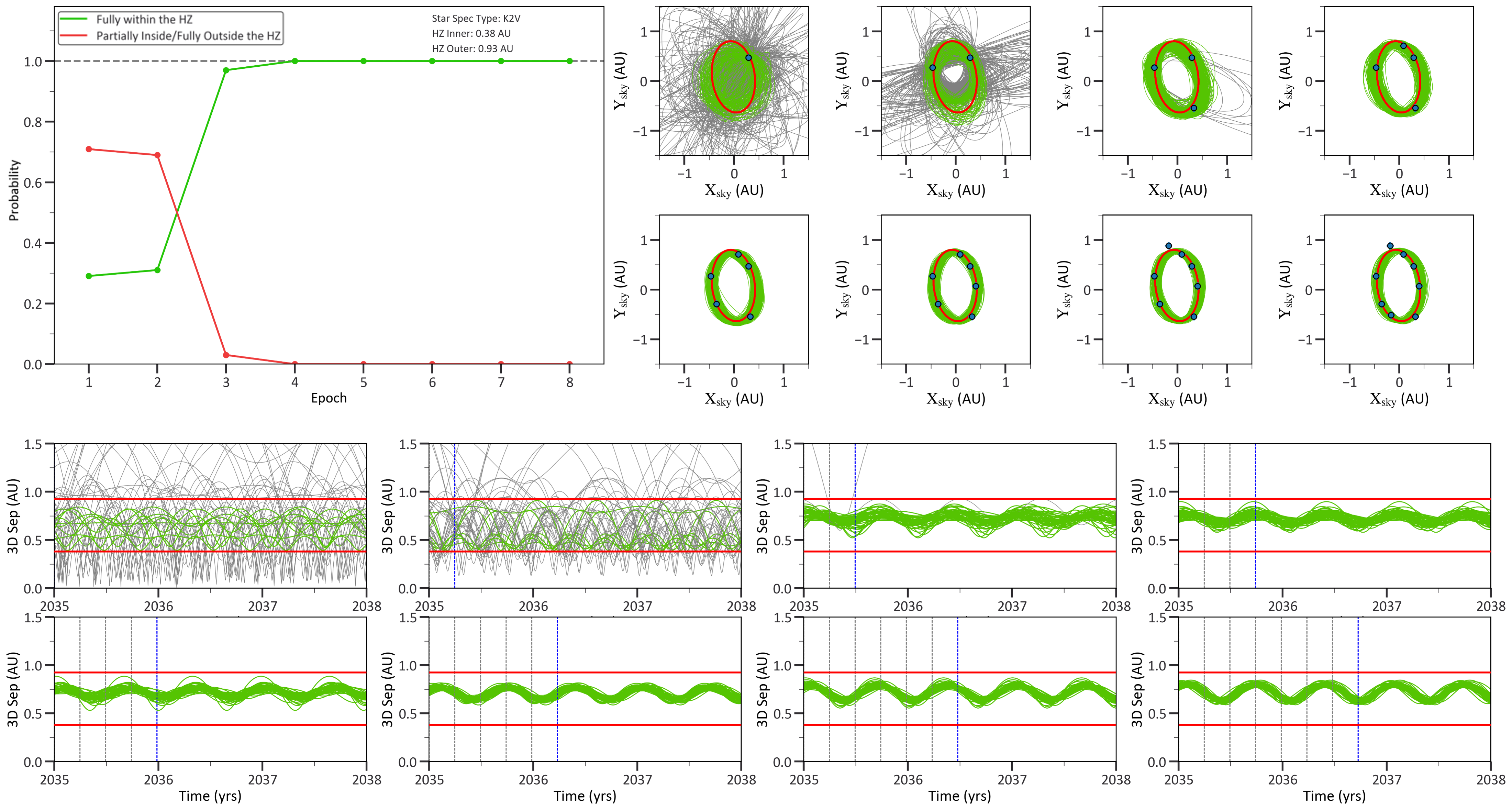}
    \caption{Confidence that a simulated Earth-like planet (a = 0.74 AU, e = 0.21) is within the star’s habitable zone (both periastron and apastron within the habitable zone (HZ) boundaries), increases with additional epochs of relative astrometry. Top left: Posterior probability on this simulated planet’s HZ status as a function of observing epoch. Green points represent the fraction of orbits from the fit posterior that are always between the inner and outer boundary of the HZ; red points either cross these boundaries, are closer to the star than the inner edge of the HZ, or are further than the outer edge of the HZ. For these simulated observations of this planet, only by epoch 4 is the probability that the planet is fully within it’s star’s HZ 100\%. The number of epochs required to confidently establish a planet as being fully within the HZ depends on the orbital geometry and the timing of the relative astrometric observations. Top right: Observed sky-projected positions of the planet (blue points), true orbit (red), and orbital fit posteriors: HZ-consistent orbits (green) and orbits extending beyond the HZ (grey). Bottom panels: Corresponding 3D star–planet separations, with optimistic HZ boundaries in red, observational epochs (dashed vertical lines) in grey, and the latest epoch highlighted in blue. Note by epoch 4, all remaining orbits are fully within the HZ boundaries}
    \label{fig:HZ Planet Fit}
    \thispagestyle{empty}
  \end{figure*}
\end{turnpage}

\begin{turnpage}               
  \begin{figure*}[p]            
    \centering
    \includegraphics[width=1.05\textheight,height=\dimexpr\textwidth-10\baselineskip\relax]{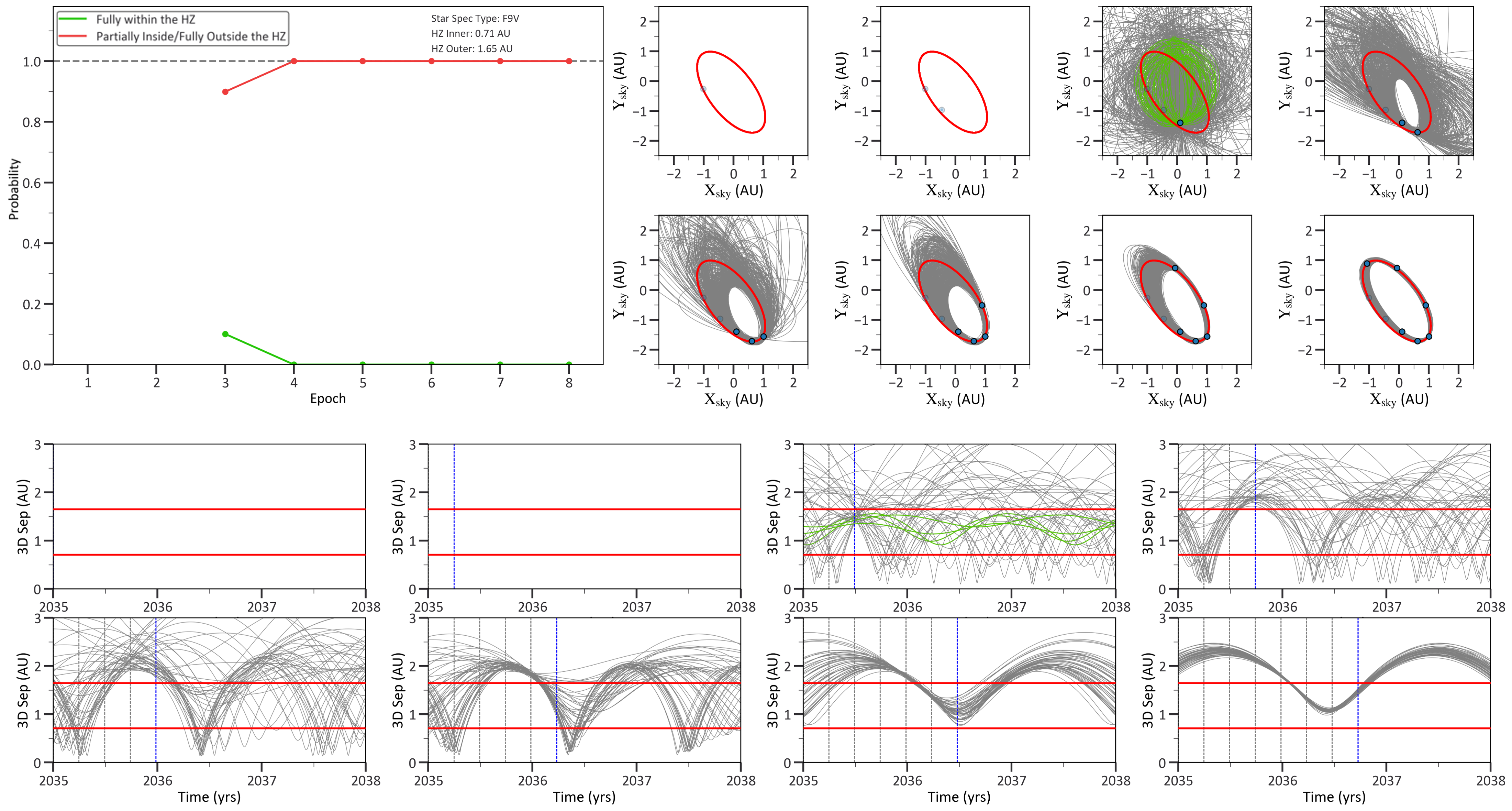}
    \caption{Same as Figure \ref{fig:HZ Planet Fit}, but for a non-HZ planet at large angular separation (a = 2.45, e = 0.43). Top left: Posterior probability on this simulated planet’s HZ status as a function of observing epoch. The large projected separations quickly rule out HZ-compatible orbits within 2–3 epochs, even with the planet undetectable in the first two epochs. Top right: Observed sky-projected positions (blue), true orbit (red), and posterior orbits: HZ-consistent (green) and inconsistent (grey). Bottom panels: 3D star–planet separations relative to optimistic HZ boundaries (red), with observational epochs (grey dashed lines) and the latest epoch in blue. By epoch 4, all remaining solutions lie outside the HZ.}
    \label{fig:Non HZ Planet Fit}
    \thispagestyle{empty}
  \end{figure*}
\end{turnpage}

Following our simulation framework for HZ planet detection and orbit recovery (see Section~\ref{subsection: HZmethod}), we simulated 98 planets around our list of target stars, identifying 29 whose orbits lay entirely within the HZ. We simulated 16 direct imaging observing epochs for each planet to assess the effectiveness of orbital characterization in confirming HZ occupancy. We illustrate this process through two representative examples: a confirmed HZ planet (Figure \ref{fig:HZ Planet Fit}) and a non-HZ planet (Figure \ref{fig:Non HZ Planet Fit}). 

In some cases, planets can be excluded from the HZ after only a few observations. Large projected separations (e.g., $>10$ AU) inherently preclude HZ occupancy, regardless of orbital orientation. In contrast, smaller projected separations may remain ambiguous, requiring additional epochs to resolve degeneracies in orbital parameters. This asymmetry is evident in the two cases shown. 

In the non-HZ case (Figure \ref{fig:Non HZ Planet Fit}), HZ-compatible orbits are ruled out entirely within 2-3 epochs. Note that the non-HZ planet is undetectable during the first two epochs. In contrast, the HZ case (Figure \ref{fig:HZ Planet Fit}) shows a steady increase in the fraction of HZ orbits, rising from $\approx 0.3$ in the first epoch to almost unity by the fourth epoch. This progression is reflected in both the orbital fits and the corresponding 3D star-planet separation plots. 

While direct imaging provides only the projected (2D) separation, the HZ status depends on the true 3D separation between a star and planet. The 3D separations for each orbital fit are shown in the lower panels of Figures \ref{fig:HZ Planet Fit} and \ref{fig:Non HZ Planet Fit}. For the HZ planet, orbital solutions gradually converge within the HZ boundaries (marked by red horizontal lines), whereas the non-HZ planet’s solutions extend far beyond them. These results suggest that while definitive HZ confirmation typically requires 4–5 epochs, large-separation non-HZ planets can often be excluded with fewer, enabling more efficient candidate screening.

\begin{figure}[ht]
\centering
\includegraphics[width=0.48\textwidth]{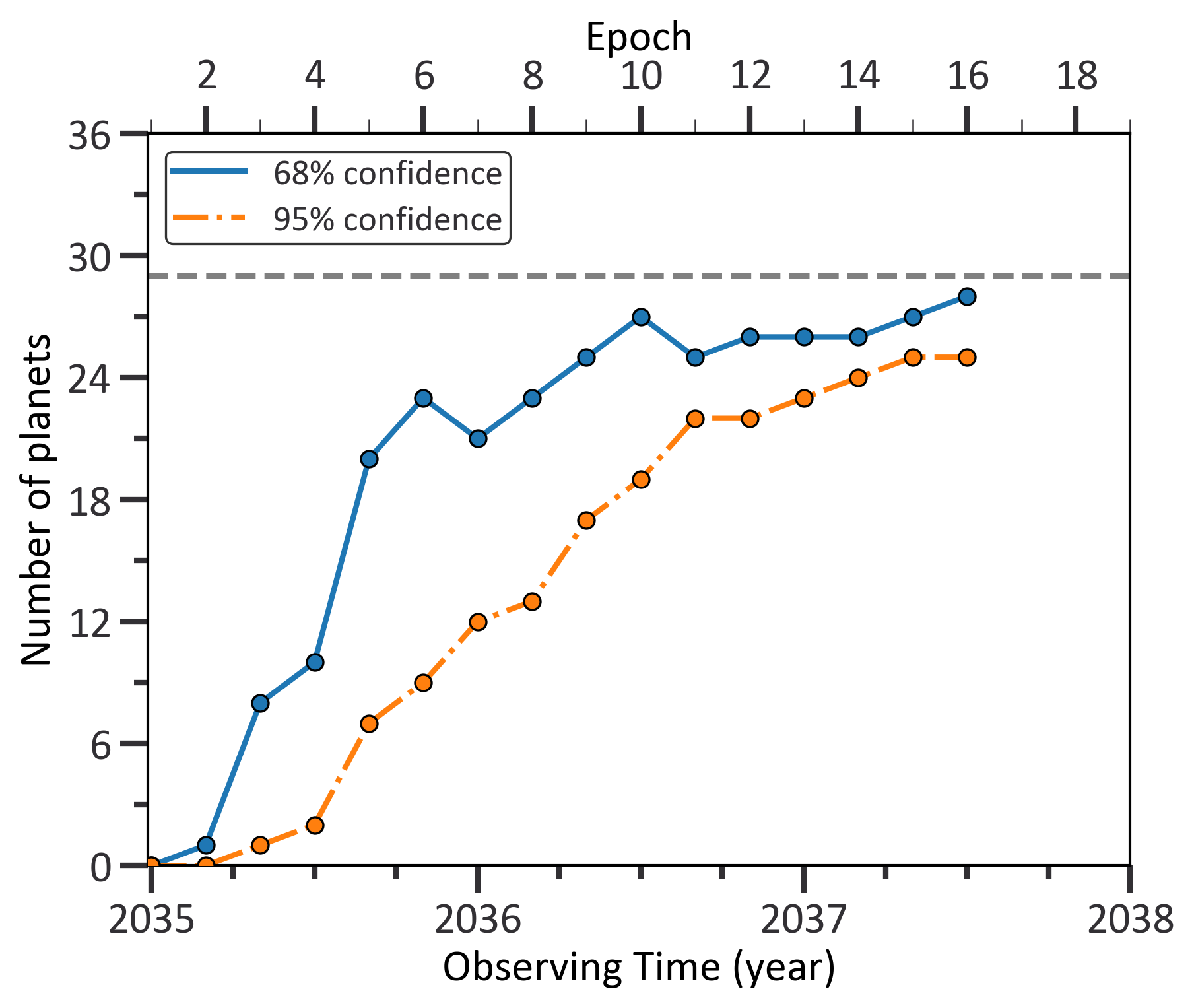}
\caption{High-confidence characterization of HZ planets requires substantially more observational time than initial identification. The plot shows the cumulative number of confirmed HZ planets versus observational epochs at 68\% confidence (solid blue) and 95\% confidence (dashed orange) levels, out of 29 true HZ planets in our 98-system sample. A planet is classified as HZ when the specified percentage of its fitted orbits lie within the optimistic HZ boundaries.}
\label{fig:Num of HZ vs epoch}
\end{figure}

To assess the efficiency of our assumed HWO architecture in characterizing HZ planets across our sample, we extended this analysis to our full sample of planets. Figure \ref{fig:Num of HZ vs epoch} shows the cumulative number of confirmed HZ planets as a function of observational epochs, using two different confidence thresholds. As expected, the 68\% confidence curve (solid blue line) - where at least 68\% of the fitted orbital posteriors fall within the HZ - rises rapidly in the early epochs, with 4-5 epochs sufficient to confirm approximately half of the true HZ planets. The rate of new confirmations slows thereafter. In contrast, the 95\% confidence curve (dashed orange line) shows markedly different behavior: minimal confirmations in early epochs followed by a sharp increase after the sixth epoch. This distinct temporal behavior reveals that while moderate-confidence HZ characterization can be achieved relatively quickly, high-confidence confirmation requires significantly more observational investment to break orbital degeneracies. Notably, even after sixteen epochs, some true HZ planets remain unconfirmed, primarily due to orbital geometries that place them outside detection limits—either from insufficient contrast or projected separations beyond the working angles.

Our results have direct implications for HWO's observing strategy and mission timeline. While rapid initial discoveries are important for demonstrating mission success, the higher confidence orbital characterizations needed for follow-up atmospheric studies require extended temporal baselines. This suggests a two-phase approach might be optimal: an initial survey phase using shorter cadence observations to identify promising candidates, followed by extended monitoring of the most interesting systems to achieve high-confidence orbital characterization. Such a strategy would maximize both early scientific return and long-term mission impact in the search for potentially habitable worlds.

% simulations, number of epochs needed, number of observations
\section{Discussion} \label{sec:discussion}

% caveats, applications, implications for mission design
% complementary results from other facilities (ELTs, ALMA, etc)
\subsection{Optimal Temporal Distribution of Astrometric Measurements}
The analysis presented in Section \ref{section: coldjupiters} demonstrates the critical importance of observation timing strategies for astrometric characterization of exoplanetary systems. Our results reveal that the temporal distribution of measurements has a profound impact on orbital parameter determination, with implications that vary significantly based on orbital period.

For planetary systems with orbital periods exceeding 100 days, we recommend implementing sparse astrometric observations distributed across the full mission duration rather than concentrating measurements within compressed timeframes. This approach leverages the extended temporal baseline to significantly enhance orbital parameter constraints, particularly for longer-period orbits where the improved phase coverage proves most beneficial for accurate orbital determination. As shown in Section \ref{section: coldjupiters}, this strategy becomes increasingly advantageous for systems with semi-major axes approaching 30 AU, where maximizing temporal coverage is essential for breaking degeneracies in orbital solutions.

\subsection{Gaia}

% add discussion on DR5, timeline
The detectability of planets around stars in the ExEP catalog using Gaia astrometry faces significant challenges, particularly for wide-separation companions. The catalog's proximity (all stars within 20 pc) would typically provide favorable conditions for astrometric detection, as closer distances enable detection of lower-mass companions. However, the brightness of the HPIC stars (V magnitudes primarily between 4-7) complicates reliable planet detection. While \cite{wallace2025} demonstrated techniques using RUWE for planet detection in Gaia DR3 as an alternative to direct astrometric measurements, the full capabilities of Gaia for planet detection will continue to evolve with future data releases.
The upcoming Gaia DR4 and DR5 releases promise significant enhancements in sensitivity for direct astrometric planet detection. The reference frame and observational approach of Gaia, as explained by \cite{brown2025}, make it particularly suited for systematic surveys of planetary companions across the sky. As detailed by \cite{brown2025}, Gaia DR4 will enable detection of massive planets with periods in the 1-5 year range. For longer-period planets, Gaia DR5 (based on the full 10.5-year mission) will be even more powerful. \cite{brown2025} shows that proper motion precision will improve by a factor of 2.8 compared to DR4, dramatically expanding the parameter space for giant planet detection to include planets with orbital periods comparable to those in our solar system.
However, we need to note that Gaia's capabilities complement HWO rather than replace it. The brightest stars in our sample ($V<6$) present challenges for Gaia measurements due to detector saturation. For these targets, which include many primary HWO targets, Gaia's uncertainties are dominated by calibration errors rather than photon noise \citep{pancino2012, refId0}.

When considering the full mission timeline, Gaia will have surveyed the sky for over a decade before HWO launches, providing valuable orbital constraints for giant planets that can optimize HWO's observing strategy. This synergy between missions -- with Gaia providing dynamical masses and orbital parameters, while HWO delivers spectroscopic characterization -- will be crucial for developing a comprehensive understanding of planetary system architectures.

\subsection{Nancy Grace Roman Space Telescope}

% it is not a survey instrument, mainly characterization
Based on the analysis by \cite{Carrion2021}, the Nancy Grace Roman Space Telescope's capability for detecting and characterizing planets through direct imaging will be determined by several key technical constraints. Roman's true strength lies in its ability to characterize individual planetary systems rather than conduct broad surveys. The coronagraph instrument (CGI), which is a technology demonstration for future space-based coronagraphs, has a bright host star requirement of $V \leq 5 \rm mag$, with the potential capability to observe stars with $V = 6-7 \rm mag$, though performance for these brighter targets remains to be determined after the technology demonstration phase.

For giant planets at wide separations (5-20 AU), the detection prospects vary significantly with orbital distance. In the optimistic scenario with $\lambda = 575 \rm nm$, Roman's inner and outer working angles (IWA = 3 $\lambda/D$, OWA = 9 $\lambda/D$) correspond to approximately 148 and 445 mas, respectively. For stars at 20 pc, planets at 5 AU (corresponding to $\sim 250$ mas) would fall within Roman's working angles, while those at 20 AU ($\sim1000$ mas) would lie beyond the OWA, making their detection impossible. 

While Roman represents a significant advance in direct imaging capabilities and will excel at the detailed characterization of individual targets, HWO will be able to overcome the limitations that prevent Roman from conducting comprehensive surveys. Missions like HWO will aim at achieving higher contrast capabilities than Roman's minimum contrast of $10^{-9}$, which will be particularly important for detecting smaller planets in the HZ. Moreover, HWO will be able to access a broader range of orbital separations, making them better suited for studying both HZ planets and those in wider orbits beyond the snow line. While Roman's capability will be primarily limited to giant planets at intermediate separations around stars of moderate brightness, HWO will be designed specifically for detecting and characterizing Earth-like planets in the HZ and will simultaneously be able to observe planets at larger separations, enabling comprehensive studies of entire planetary systems and their architectures through systematic surveys rather than targeted observations.

\subsection{Needed Precursor RV Observations}

The successful characterization of giant planets with HWO will require substantial precursor radial velocity observations \citep{kane2024e}. Our analysis indicates that approximately 40 RV measurements per target, with a precision of $\sim1$ m/s, are necessary to adequately constrain the orbital parameters and identify optimal timing windows for direct imaging observations. This requirement stems from the need to precisely determine orbital periods, phases, and approximate inclinations of potential targets before investing valuable telescope time. We note that such precursor RV work has already commenced, such as the evaluation of archival RV data by \cite{Laliotis2023,Harada2024,Howard2016}.

RV observations of this frequency help establish reliable orbital ephemerides, enabling HWO to schedule its high-contrast imaging at phases when planets are expected to be at their maximum projected separation from their host stars. For giant planets with periods between 15-85 years, these RV measurements should ideally span multiple years to capture sufficient orbital motion. While this represents a significant investment of ground-based telescope time, it is essential for maximizing the scientific return of HWO's limited observing window.

However, it is important to note that not all habitable-zone planets discovered by HWO will be suitable for precise RV follow-up to measure their masses. The RV signal induced by an Earth-mass planet in the HZ is at the cm/s level for Sun-like stars, pushing the limits of foreseeable precursor RV capabilities. Factors like the star's brightness, spectral type, activity level, and the orbital separation of the planet can make some HWO targets much more challenging for precursor RV mass measurements.

The combination of comprehensive RV monitoring with HWO's precise astrometry will provide the most complete orbital solutions, particularly for systems with periods shorter than the mission lifetime. Our simulations demonstrate that this multi-technique approach, when properly timed using RV-derived orbital predictions, yields the most precise constraints on planetary masses and orbital architectures. 

\section{Conclusions}\label{sec:conclusions}
Our comprehensive analysis of HWO's capabilities for joint detection and characterization of habitable worlds and cold giants yields several key findings. First, we recommend a coronagraph with an OWA of at least 1440 mas to achieve 80-90\% detection completeness for cold giant planets across our stellar sample while maintaining technical feasibility. This specification ensures that the majority of planets in target systems will be visible for a significant portion of their orbits, enabling accurate orbital characterization. Second, our simulations demonstrate that 6-8 astrometric measurements distributed across the five-year mission timeline provide optimal constraints on orbital parameters for planets with periods under 100 years. For longer-period planets ($>100$ years), even an optimized temporal baseline provides insufficient coverage for reliable orbital characterization.

We find that combining precursor RV measurements with HWO's astrometric capabilities significantly enhances orbital parameter precision, particularly for inclination and eccentricity determination. This synergy between ground-based RV monitoring and space-based direct imaging will be essential for developing a comprehensive understanding of planetary system architectures. Furthermore, our results indicate that while moderate-confidence HZ characterization can be achieved with 4-5 epochs, high-confidence confirmation suitable for follow-up atmospheric studies requires a more extended observational timeline.

The successful characterization of giant planets with HWO will require substantial precursor radial velocity observations, with approximately 40 RV measurements per target recommended to adequately constrain orbital parameters. The ability to simultaneously characterize both HZ planets and cold giants will provide crucial insights into the dynamical processes that influence habitability, making this dual-detection capability an essential aspect of HWO's mission design. These findings collectively establish baseline observational requirements that will maximize scientific return in the search for potentially habitable worlds.

\section*{Acknowledgments}
\begin{acknowledgments}
This analysis was performed as part of the HWO Demographics \& Architectures Working Group steering committee.
We thank the anonymous referee for the insightful comments. We thank Diana Dragomir, William Roberson, Peter Wheatley, and Rob Wittenmyer for endorsing this science case. S.S. thanks Courtney Dressing, Andy Casey, Matthew Kenworthy, Maxwell Millar-Blanchaer, and Will Farr for the valuable discussions. S.S. also acknowledges a graduate fellowship at the Kavli Institute for Theoretical Physics, during which this research was completed. S.S. thanks the LSSTC Data Science Fellowship Program, funded by LSSTC, NSF Cybertraining Grant number 1829740, the Brinson Foundation, and the Moore Foundation; her participation in the program has benefited this work. M.R. acknowledges support from Heising-Simons Foundation grants $\#$2021-2802 and $\#$2023-4478, NASA Exoplanets Research Program NNH23ZDA001N-XRP (grant $\#$80NSSC24K0153), and the National Geographic Society. CKH acknowledges support from the National Science Foundation Graduate Research Fellowship Program under Grant No. DGE 2146752. T.D. acknowledges support from the McDonnell Center for the Space Sciences at Washington University in St. Louis. Y.H. was supported by the Jet Propulsion Laboratory, California Institute of Technology, under a contract with the National Aeronautics and Space Administration (80NM0018D0004). P.J.A. acknowledges support from NASA TCAN award 80NSSC19K0639, and from award 644616 from the Simons Foundation.

\end{acknowledgments}


\vspace{5mm}


\software{\texttt{numpy} \citep{numpy2020}, \texttt{matplotlib} \citep{matplotlib2007}, \texttt{astropy} \citep{astropy:2013, astropy:2018, astropy:2022}, \texttt{octofitter} \citep{Thompson2023}, \texttt{starry} \citep{Luger2019},
\texttt{orbitize}
\citep{Blunt2019}.}




\bibliography{hwo-dynamics}{}
\bibliographystyle{aasjournal}

\end{document}